\documentclass[useAMS,usenatbib,usegraphicx]{mn2e} 

\usepackage[intlimits]{amsmath}
\usepackage{amssymb}
\usepackage{graphicx}
\def\ang{$\rm{\AA}\;$}                           

\def\Msolar{\mbox{$\rm{M}_{\normalsize\odot}$}}
\def\gt{$>$}

\def\la{\mathrel{\hbox{\rlap{\hbox{\lower4pt\hbox{$\sim$}}}\hbox{$<$}}}}
\def\ga{\mathrel{\hbox{\rlap{\hbox{\lower4pt\hbox{$\sim$}}}\hbox{$>$}}}}
\def\arcmin{\hbox{$^\prime$}}
\def\arcsec{\hbox{$^{\prime\prime}$}}
\def\slantfrac#1#2{\hbox{$\,^#1\!/_#2$}}

\def\onequarter{\slantfrac{1}{4}}

\def\br{$B\!-\!R\;$}               
\def\nodata{\multicolumn{1}{c}{$\cdots$}}
\def\ion#1#2{$\rm{#1}$\;${\small\rm\@Roman{#2}}$\relax}

\def\pcm2{\rm{cm}^{-2}}
\def\logcoldens{$\rm{log_{10}\left(N_{HI}\left(cm^{-2}\right)\right)}$}
\def\hMpc{h$^{-1}\;$Mpc}

\def\H0{$\rm{H_{0}}$}
\def\kms{$\rm{km\;s^{-1}}$}

\def\hMpc{$\rm{h^{-1}\;Mpc}$}
\def\hkpcsev{$\rm{h^{-1}_{70}\;kpc}$}
\def\hMpcsev{$\rm{h^{-1}_{70}\;Mpc}$}
\def\twopcf{$\xi_{\rm{AG}}$}
\def\acf{$\xi_{\rm{GG}}$}

\def\lya{Ly-$\alpha$}
\def\lyb{Ly-$\beta$}
\def\kms{km s$^{-1}$}
\def\gtrsim{\mathrel{\hbox{\rlap{\hbox{\lower4pt\hbox{$\sim$}}}\hbox{$>$}}}}
\def\lesssim{\mathrel{\hbox{\rlap{\hbox{\lower4pt\hbox{$\sim$}}}\hbox{$<$}}}}
\newcommand{\CIV}{\hbox{{\rm C}{\sc \,iv}}}

\newcommand{\FeII}{\hbox{{\rm Fe}{\sc \,ii}}}

\newcommand{\MgII}{\hbox{{\rm Mg}{\sc \,ii}}}

\newcommand{\OVI}{\hbox{{\rm O}{\sc \,vi}}}

\title[The Association Between Gas and Galaxies]{The association
between gas and galaxies III: The Cross-correlation of Galaxies$^{\dagger}$
and \lya\ Absorbers$^{\ddagger}$ at z$\sim$1}
\author[Shone, Morris, Crighton \& Wilman]{\parbox[h]{160mm}{Allen M. Shone$^{\star1}$ Simon L. Morris$^{1}$ 
Neil Crighton$^{1}$ Richard J. Wilman$^{2}$}
\vspace{6pt} \\
$^1$Department of Physics, University of Durham, South Road, Durham, DH1 3AJ, UK.\\
$^2$Centre for Astrophysics \& Supercomputing, Swinburne University of Technology, Hawthorn, Victoria 3122, Australia.}

\begin{document}

\date{Accepted 0000 January 00. Received 0000 January 00; in original form 0000 January 00}

\pagerange{\pageref{firstpage}--\pageref{lastpage}} \pubyear{2009}
\maketitle
\label{firstpage}
\begin{abstract}
We have measured the 2D 2-point correlation function, \twopcf,
between low column density \lya\ absorbers and galaxies at a redshift
$\rm{z}\sim1$. We measured \lya\ absorbers between redshifts ${\rm
z}=0.68\rightarrow1.51$ over a total redshift path length of
$\Delta{\rm z}=1.08$ from HST STIS E230M absorption
spectra towards the quasars HE 1122-1648 ($\rm{z}=2.4$) and PKS
1127-145 ($\rm{z}=1.187$).  The column density of the \lya\
absorbers ranged from $13.2\leq$\logcoldens$\leq17.4$, with a median
column density of \logcoldens$=14.0$.  A total of 193 galaxy redshifts
within the surrounding 6.8\arcmin$\;\times\;$5.7\arcmin\ field of view
of both quasars were identified in a ${\rm R}$ magnitude limited
survey ($21.5\leq \rm{R}_{\rm{Vega}}\leq24.5$) using the FORS2
spectrograph at the VLT, of which 95 were higher than 
the minimum redshift of $\rm{z}=0.68$ to be used in the correlation
function. A $3\sigma$ upper-limit of \twopcf$=2.8$ was
found when 145 \lya\ absorber-galaxy pairs were binned in redshift
space, in a bin of size $\Delta\sigma=1.0$, $\Delta\pi=2.0$ \hMpcsev\
along the projected separation and line of sight distances
respectively.

The upper-limit in the cross-correlation was found to be $5.4\sigma$
lower than the central peak in the galaxy auto-correlation within the same
redshift range, \acf, which was in our data equal to $10.7\pm1.4$.
Thus we have shown for the first time that the
clustering between low column density absorbers and galaxies at
a redshift of 1 is weaker than that between galaxies at the same
redshift.
\end{abstract}

\begin{keywords} galaxies -- intergalactic medium, galaxies -- quasars:
absorption lines, galaxies -- galaxies:  haloes.
\end{keywords}

\footnotetext{$^{\dagger}$Based on observations collected at the
  European Organisation for Astronomical Research in the Southern
  Hemisphere, Chile. ESO programme number 076.A-0312.}
  \footnotetext{$^{\ddagger}$Based on observations made with the
  NASA/ESA Hubble Space Telescope, and obtained from the Hubble Legacy
  Archive, which is a collaboration between the Space Telescope
  Science Institute (STScI/NASA), the Space Telescope European
  Coordinating Facility (ST-ECF/ESA) and the Canadian Astronomy Data
  Centre (CADC/NRC/CSA). Additional data based on observations
  from the UVES spectrograph was collected at the European
  Organisation for Astronomical Research in the Southern Hemisphere,
  Chile. ESO programme numbers 067.A-0567(A),068.A-0570(A) and
  069.A-0371(A).}
\footnotetext{$^{\star}$E-mail: a.m.shone@durham.ac.uk}

\section{Introduction} \label{sec-intro}
This paper is the third in a series that investigates the extent to
which \lya\ absorbers in the inter-galactic medium (IGM) are
associated with galaxies.

At the current epoch only a minor fraction of the baryons in the
universe exist within galaxies. Stars, galaxies and other stellar
remnants make up 6\% of the baryonic matter density of the universe,
$\Omega_{\rm b}$, while within the galaxies cold ($\rm{T}\lesssim10^4$
K), condensed gas contributes a further $\sim1.7\%$
\citep{Bregman2007}. Therefore at least 90\% of baryons in the
universe lie beyond the galaxies, in the IGM.

Part of this remaining fraction, which is only $10-100$ times the mean
cosmic density, is more susceptible to shock heating. Shock waves that
are caused by gravitational collapse travel along the filamentary
structure heating the gas to $\rm{T}=10^5-10^7$ K. These baryons form
the diffuse warm-hot intergalactic medium (WHIM)
\citep{CenOstriker1999,Fukugita1998}. The detection and study of the
WHIM using metals detected in ultra-violet \citep{Verner1994,Tripp2000a,Danforth2005,Thom2008} 
and x-ray spectroscopy \citep{Nicastro2002,Fang2006}
is a major science motivation
for the future space instruments XEUS (the X-Ray Evolving Universe
Spectrometer) and the new COS (Cosmic Origins Spectrograph) on the
Hubble-Space Telescope (HST).

A third of the gas that remains is probably located in the diffuse IGM
at temperatures $\rm{T}\sim10^4-10^5$ K
\citep{Sembach2004,Lehner2007,Danforth2008}.  The neutral fraction of
this ionised plasma is observed in quasar absorption spectra. Under
the influence of dark matter this gas is thought to have collapsed to
form clouds that are part of a filamentary network that traces the
layout of the galaxies.  Two original opposing models questioned
whether this gas is correlated with galaxies but is part of the giant
cosmic web \citep{Morris1993}, or whether the clouds are discrete and
isolated with the gas contained within galaxy haloes
\citep{Lanzetta1995}.

Even though opinion is siding with the former there is probably no
definitive answer to this question, as the typical location of these
clouds is expected to be a function of their column density. 
At \logcoldens$\sim21$, the density of gas in an absorber
would be comparable to that in the outer regions of a galaxy.
Therefore at a small separation it would be pointless to draw a line
where a galaxy terminates and the absorber begins.  In the low
\citep{Rao2003,Zwaan2005} and intermediate redshift universe 
\citep{Brun1997,Chen2003} clouds with a column density 
\logcoldens$\gtrsim20.3$, 
(the damped \lya\ absorbers systems) have been 
been shown in imaging and spectroscopy to be associated 
with galaxies.  Meanwhile those clouds with a column
density several decades less dense, \logcoldens$\leq14$, are more
likely to make up the filaments or lie in the voids of the IGM.
The absorber-galaxy association is also dependent on the epoch, for 
example because of the decrease in the number
density of \lya\ lines since redshift $\rm{z}\sim1.7$
\citep{Sargent1980,Bahcall1991,Janknecht2006}.

\subsection{The Issues to Resolve and Investigations so far}
The 2D 2-point cross correlation between \lya\ absorber and galaxy positions is a
powerful way to quantitatively measure the relationship between galaxies
and the IGM. It is hoped that these measurements will
provide an insight into the degree of association
between galaxies and the surrounding neutral gas. The 
degree of correlation is thought to depend 
on the redshift, the column density of the absorbing cloud, 
the galaxy magnitude, the presence of a galactic wind, 
and the line-of-sight distance and projected 
separation (distances $\pi$ and $\sigma$) 
between the galaxy and the absorber. 

A common way to analyse the \lya\ absorber-galaxy 
cross-correlation is to contrast the results with the
galaxy auto-correlation. If these values are comparable
or the cross-correlation is greater 
at small separation, then the \lya\ absorbers
could be claimed to be more of a feature of the 
galactic halo rather than the IGM. 

We now give a short overview of such measurements that
have been made over different redshifts.

\subsubsection{Observations at Low Redshift}
\citet{Weber2006} (hereafter RW06) studied the HI-galaxy correlation
at redshifts $\rm{z}\sim0$.  5317 gas rich galaxies that were
detected in the HIPASS galaxy survey \citep{Zwaan2003} were correlated
with 129 absorbers from 27 lines of sight.  These spectra were
gathered from STIS and GHRS data and all absorbers had
\logcoldens$=12.41-14.81$.

It was found that at small separations there is a
strong correlation between absorbers and gas rich galaxies that is
similar in strength to the galaxy auto-correlation function.  This
supports the idea of matter clustered within the vicinity of galaxies
(within the nodes of the cosmic web) be they absorbing clouds or other
galaxies. However, unlike the galaxy auto-correlation that decays
radially along the $\sigma$ and $\pi$ direction, \twopcf\ remained
high out to 10 \hMpc\ or 1000 \kms\ along the line of sight. The
proposed origin of this was the accretion of gas. In this scenario the
gas falls in from the more diffuse IGM, drains along filaments and
falls into a galactic halo causing an elongation in the redshift space
correlation along the line of sight.  The gas rich galaxies that were
detected using 21 cm radiation also tended to have a low mass
($\rm{M}\sim10^{8}$ \Msolar), and it is galaxies such as these that
\citet{Keres2005} claims are dominated by this method of accretion.

\subsubsection{A simulation of \twopcf\ at low redshift}
The results of RW06 were compared with simulations in
\citet{Pierleoni2008}. Three simulations, each with a different galaxy
wind model were run to a redshift of zero. Either no winds, strong
winds or extremely strong winds were present.
 
Spectra were generated from 999 sightlines that sampled the
simulation. The \lya\ absorbers that had a column density in
the range \logcoldens$=12.41-14.81$ were then correlated with
$\sim5000$ galaxies that resided in haloes of mass
$\rm{M}=8\times10^{10}-10^{13.5}$ \Msolar.   
However, contrary to RW06, \citet{Pierleoni2008} found the galaxy
auto-correlation was predicted to be stronger than the absorber-galaxy
cross-correlation.

The correlation function was recalculated using only 27 sightlines to
match the sample from RW06, and the sample noise increased. Bootstrap
errors showed a difference between \twopcf\ and \acf\ at only a
$1\sigma$ significance level.  \citet{Pierleoni2008} suggest that it
is this sparse sampling that make the cross and auto-correlation
functions indistinguishable. They also attribute the large
``finger-of-god'' observed in Figure 3 of RW06 to this effect.

\subsubsection{Observations at low to intermediate redshift}
\begin{enumerate}
\item{{\it\citet{ChenMulchaey2009,Chen2005}}\\
\\
\citet{Chen2005} studied the sightline towards PKS 0405-123, and
compared the positions of 112 \lya\ absorbers from STIS echelle
spectra with 482 galaxies within the redshift range $\rm{z}=0.1\rightarrow0.5$. 
These were collected using a magnitude
limited survey down to $R\leq20$.

The most striking result found by \citet{Chen2005} was a difference in
the level of correlation that depended on whether the galaxy in
question was dominated by emission or absorption spectral lines.  The
correlation function between galaxies that were dominated by emission,
and \lya\ absorbers that had \logcoldens$\geq14.0$, was found to be
similar to the galaxy auto-correlation. No significant
cross-correlation was seen for those galaxies dominated by absorption.

\citet{ChenMulchaey2009} then expanded on this result with two further
lines of sight towards quasars HE 0226-4110 and PG 1216+069. It was
found that within a projected separation of 1 $\rm{h_{100}^{-1}\;Mpc}$
\lya\ absorbers with \logcoldens$\ge14$ had a similar
cross-correlation with galaxies as the auto-correlation of emission
line dominated galaxies. This was $\sim6$ times weaker than the
self-clustering of absorption line dominated galaxies. 
They suggest that the cross-correlation was similar between 
emission dominated galaxies and \lya\ absorbers because both
these occupy the same halo system. 

The cross-correlation between any type of galaxy and \lya\
absorber with \logcoldens$\le13.5$ was observed to be very weak.} 
\\
\item{{\it{Papers I and II}}\\ 
\\
This paper is an extension to the investigation first conducted in
paper I, \citet{M_J_2006} (hereafter MJ06). Here the location of 636
galaxies, with redshifts determined using CFHT spectroscopy, were
compared with 381 \lya\ absorbers found in the UV spectra of quasars
taken with the Faint Object Spectrograph during the Hubble Key
Project.  It was found that, excluding the high column density
systems (\logcoldens$\geq17$), 
absorbers are correlated with galaxies at redshifts of
$\rm{z}\leq0.8$ and that this correlation is weaker than the
corresponding galaxy auto-correlation.

Paper II \citep{Wilman2007} (hereafter W07) using the MJ06 dataset
computed the 2-point correlation function along the line-of-sight
($\pi$) and projected separation ($\sigma$). A conclusion drawn from
this work suggested that at small co-moving separations of $\sigma<
0.4$, $\pi<2\;\rm{h}^{-1}$Mpc, the correlation function increases
marginally as the column density increases from \logcoldens$=13-16$,
then there is suggestion of a sharp increase in correlation above
\logcoldens$\geq17$.  This was proposed as tentative evidence for the
column density at which absorbers become a part of the galactic
halo.

It was argued in W07 that a ``finger-of-god'' may also be caused
by a large anisotropy, for example a quasar line of sight that runs
parallel and close to to a large filament. However \citet{Pierleoni2008}
ruled out a ``finger-of-god'' being caused by this geometric effect in
their simulation by showing that the \twopcf\ result remained the same
no matter which orientation the sightlines had been directed.
Instead a ``finger-of-god'' was caused if a sightline passed 
through a virialised system of gas and a paired galaxy.}
\end{enumerate}

\subsubsection{Observations at high redshift}
In order to observe the IGM-galaxy relationship at high redshift
(where the \lya\ absorption has now reached optical wavelengths)
\citet{Adelberger2003} correlated the \lya\ lines from 8 quasar HIRES
or ESI spectra with 431 Lyman-break-galaxies.

It was expected that greater concentrations of gas would be located
near to the galaxies. This was found to be so as gas with a greater
optical depth than the mean density was found to lie within $\sim10$
\hMpc\ of the nearest galaxy.  However, evidence arose of an
anti-correlation within $\sim0.5$ \hMpc.  It was hypothesised that
this was caused by super-winds that, travelling at $\sim600$ \kms\
\citep{Pettini2001}, clear the immediate vicinity of the galaxies and
heat the surroundings, decreasing the neutral fraction of HI and so
increasing the transmission of QSO flux.

With a larger data set of 23 sightlines and 1044 UV selected galaxies
\citet{Adelberger2005} found that evidence for an anti-correlation at
small separation significantly decreased. Only in a third of the cases
was weaker absorption now detected around galaxies.

\subsection{The motivation for this work}
No results have been obtained for the redshift window
$\rm{z}\sim1$. The main reason for this is the difficulty of acquiring
a sufficient number of galaxy redshifts at this epoch that are within
the same fields of view as quasars for which there is high resolution
UV echelle spectral data. The aim of this investigation was to start
to bridge this gap.

Thus we hope to shed light on the \lya\ absorber-galaxy correlation
at redshifts $\rm{z}\sim1$. Then by comparison with
the other works mentioned above determine whether the correlation
evolves with redshift.

The structure of this paper will be as follows. The data reduction
for the FORS2 galaxy survey and absorbers from the HST
STIS E230M spectra will be described in Sections 2 and 3 respectively.
In Section 4 the basics of the 2D 2-point cross correlation function
used will be outlined.  Results are given in Section 5 and we present
our discussion and conclusions in Section 6.

Throughout the paper we have adopted the cosmological parameters
$\Omega_{\rm{M}}=0.3$, $\Omega_{\rm{\Lambda}}=0.7$, $\rm{h_{70}}=0.7$ and 
${\rm H}_0=100\rm{h}_{70}$ km~s$^{-1}$~Mpc$^{-1}$.

\section{The Galaxy Survey}
Galaxy surveys in two $6.8\arcmin \times 5.7\arcmin$ regions around
  the quasars HE 1122-1648 ($z=2.40$) and PKS 1127-145 ($z=1.187$)
  \citep{Carswell2002,Bechtold2002} were carried out using the FORS2
  spectrograph at the VLT on the 20-22 February 2006
  \citep{Fors2}. The total wavelength range of the
  absorbers to be used in the archived HST STIS data was between
  2050-3100 \AA. This means we could potentially detect Ly$\alpha$
  absorbers between the redshifts of 0.68 and 1.55. It
  was our aim to observe galaxies within a similar redshift
  window. This was done by selecting galaxies based on a magnitude
  limit ($21.5\leq\rm{m}_R\leq24.5$) and using FORS2 to detect
  [O II] emission appearing in the 5600-11000 \ang bandpass. It is
  thought these [O II] emitters, indicative of star-forming galaxies,
  with their systems of feedback, outflows and photoionisation are
  responsible for most of the complexity in the gas-galaxy
  relationship at z$\sim1$.

\subsection{Pre-imaging and mask generation} 
Prior to the Multi-Object-Spectroscopy (MOS) observing run pre-imaging
was taken in service mode of the two quasar fields using the Bessel
R\_SPECIAL+76 filter.  Photometric conditions with good ``seeing''
($\leq0.8$) was requested for each 10 minute exposure in order to
attain accurate astrometry and photometry when selecting faint
($R\leq24.5$) galaxies.

Table 1 shows the areas of sky that were included.

\begin{table}
\label{tab:areaofsky}
\caption{The quasar field coordinates}
\begin{tabular}{lll} \hline
\hline
Quasar &  RA range & DEC range \\
 & (J2000) & (J2000) \\
\hline
PKS 1127-145 & $11\; 30\; 21.57:11\; 29\; 52.50$ 
& $-14\; 53\; 34.84:-14\; 45\;  54.81$ \\
HE 1122-1648 & $11\; 24\; 57.47:11\; 24\; 28.07$ 
& $-17\; 09\; 25.71:-17\; 01\;  45.68$ \\ \hline
\end{tabular} \\
\end{table}

These images were reduced using the FORS2 imaging pipeline at ESO
headquarters in Garching as part of the routine service mode.
Calibration data had also been taken at the same time, including
bias frames, sky flats and photometric standard images. Using this
calibration data the raw images were bias subtracted, flat-fielded and
cosmic rays were removed.

We performed the final stages needed to reduce the science images 
locally using the standard IMUTIL tasks within IRAF.\footnote{IRAF is
distributed by the National Optical Astronomy Observatory, which is
operated by the Association of Universities for Research in Astronomy
(AURA) under cooperative agreement with the National Science
Foundation.}  The overscan region of each frame was cropped and then
images taken from chips 1 and 2 were joined together using IMTILE.

The software Source Extractor (v2.4) \citep{SExtractor} was then run
in order to compile a catalogue of all the possible targets for
follow-up spectroscopy.

SExtractor was used to calculate the photometry of every target and
classify each target according to whether it is a star or galaxy.  The
relevant parameters were MAG AUTO, an estimate of the apparent
magnitude enclosed within an ellipse that encompasses 90\% of the the
flux, and CLASS STAR.

The magnitude zero point was determined by measuring the total counts of
exposed standard stars within the image foreground and comparing these
flux values with that of their magnitude in the USNO-A2.0 star
catalogue.

\begin{figure}  
\centering  
\includegraphics[scale=0.48]{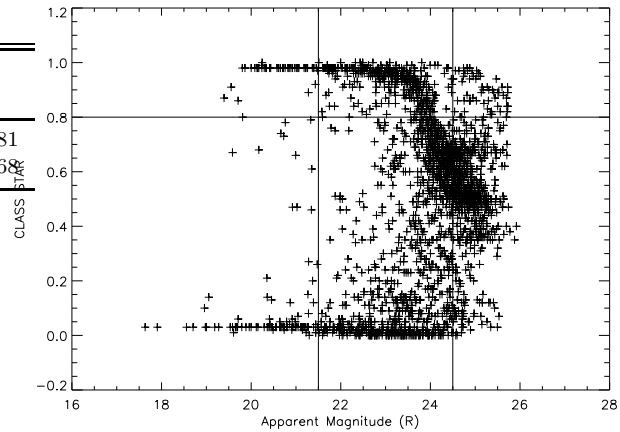} 
\caption{This horseshoe-shaped plot of the CLASS STAR classification
against magnitude allowed cuts to be made so that galaxies within the
enclosed rectangle could be included in the galaxy catalogue to be
sent to the mask making software. Those selected had $21.5\le {\rm{m}_R \le
24.5}$ and $0 \le\; {\rm CLASS STAR}\;\le 0.8$.}
\label{fig:galstar}
\end{figure}

Figure \ref{fig:galstar} shows a plot of CLASS STAR against apparent
magnitude (m$_{R}$) for all the 2666 objects extracted from the field
of HE 1122-1648.  One can tell that the reliability of CLASS STAR
decreased as the targets become fainter. A cut of CLASS STAR $\leq
0.8$ was made to select galaxy candidates
\citep{Classstar08_SExtractor}.

This plot also gave an indication as to what would be an effective
faint magnitude cut-off.  To avoid those targets where CLASS STAR was
ambiguous a faint cut of m$_R\leq24.5$ was chosen. The minimum
brightness was also set by what could be successfully observed within the allotted
exposure time. To remove those foreground galaxies likely to
have redshifts too low to correlate with the HI absorbers, a bright
cut-off of m$_R\geq 21.5$ was employed.

\subsubsection{Making the masks}
The final catalogues consisted of 1062 and 1122 objects in the fields
of HE 1122-1648 and PKS 1127-145 respectively.  The lists were then
passed to the FIMS software for the FORS2 instrument that designs
custom masks for the mask-exchange-unit (MXU).

We designed rectangular slits 1\arcsec\ by 8\arcsec. This
meant that all targets took up no more than $\onequarter$ of the
length of the slit, allowing ample space for sampling the sky so
that it could be subtracted later.  The slit sizes and the galaxy
density meant that up to 50 slits could be made per mask.

When investigating 2D 2-point correlation functions it is important
that the slits are assigned randomly so as not to bias a particular
absorber-galaxy/galaxy-galaxy angular separation.  Therefore a random
slit allocation was used in FIMS to ensure a random distribution of
galaxies was selected when designing each mask.

It was also important to obtain more than one mask per field-of-view
to allow sampling of close pairs.

The catalogue was split between those galaxies with $21.5\le$ m$_{R}
\le23.5$ and $23.5\le$m$_{R} \le 24.5$ so that separate ``bright'' and
``faint'' masks would be used. They could then be exposed for
different periods of time.  This allowed us to optimise the exposure
times used to reach a given signal-to-noise ratio for each mask -
shorter exposure times for bright masks, and longer times for faint
masks. Altogether 22 masks were designed with each mask containing
approximately 40 slits.  There were 11 for each field-of-view,
including 5 that were assigned bright objects and 6 that had slits
over galaxies with $23.5\le$m$_{R} \le24.5$.

\subsection{Details of the observing run}
The FORS2 MXU was used in visitor mode with the R\_SPECIAL+76 filter
and the 200I+28 grism. This
grism gave a central wavelength of 7450 \ang with a spectral range of
$5600\rightarrow 11000$ \ang.  The dispersion was $2.43
\;\rm{\AA}\;\rm{pixel^{-1}}$ with a resolution
$\frac{\lambda}{\Delta\lambda} = 380$ \citep{Fors2man}.

 The plan for the first night of observation (MJD 53787) was to
concentrate on PKS 1127-145 with 5 ``faint'' masks and 5
``bright''. Very poor `seeing' conditions ($\sim1.4\arcsec$) and high
winds hampered the first night of observing and it was found that
``faint'' masks could not be observed in these conditions.

For the second night (MJD 53788) we concentrated on the ``bright''
masks only from both fields-of-view.  The weather improved
significantly together with the `seeing' which remained $\le1''$.
During the first part of the second night there was a
target-of-opportunity (TOO) that interrupted the first observation.
Therefore the 1122 ``Bright'' 1 mask was observed again in
service-mode. Two exposures of 1750 seconds each were carried out in
good conditions for each bright mask.  The data we were able to
observe over three nights is summarised in Table 2.

\begin{table}
\label{tab:observations}
\caption{The galactic spectroscopy that was carried out}
\begin{tabular}{cccc} \hline
\hline
Mask & No. of Exposures & Mean & Comments \\
name & and time (s) & 'Seeing' & \\
\hline
 1127 ``Bright'' 1 & 3 x 650 & 1.7\arcsec\  &\nodata   \\
 1127 ``Faint''  1 & 2 x 1650 &1.4\arcsec\  & No continuum\\ 
 1127 ``Faint''  2 & 2 x 1650 &1.0\arcsec\  & Few sources\\
 1127 ``Bright'' 2 & 3 x 650 & 1.0\arcsec\  &\nodata  \\ 
 1127 ``Bright'' 3 & 3 x 650 & 0.80\arcsec\ &\nodata  \\
 1127 ``Bright'' 4 & 3 x 650 & 0.80\arcsec\ &\nodata  \\
 1127 ``Bright'' 5 & 1 x 650 & 1.7\arcsec\  &\nodata \\
\hline		       		    
 1122 ``Bright'' 1 & 2 x 1750 & 1.0\arcsec\ &  TOO \\
 1122 ``Bright'' 2 & 2 x 1750 & 0.75\arcsec &\nodata  \\
 1127 ``Bright'' 1 & 2 x 1750 & 0.65\arcsec &\nodata  \\
 1127 ``Bright'' 2 & 1 x 1750 & 0.50\arcsec &\nodata  \\
 1127 ``Bright'' 3 & 1 x 1750 & 0.65\arcsec &\nodata  \\
 1127 ``Bright'' 4 & 1 x 1750 & 0.65\arcsec &\nodata  \\
 1127 ``Bright'' 5 & 2 x 1350 & 0.65\arcsec &\nodata  \\
 1122 ``Bright'' 3 & 2 x 1350 & 0.65\arcsec &\nodata  \\
 1122 ``Bright'' 4 & 2 x 1200 & 0.70\arcsec &\nodata   \\
\hline					   
 1122 ``Bright'' 1 & 2 x 1750 & 0.65\arcsec &\nodata \\ \hline
\end{tabular}
\end{table}

\subsection{Data reduction of the FORS2 data set}
At the time of reduction no ESO pipeline existed for the automatic
reduction of FORS2 MXU data, so the extraction of galaxy spectra was
mainly performed using IRAF with predominant use of the IMUTIL and
NOAO packages.

The science frames were stacked
to remove contamination by cosmic rays. The majority of
these were removed using COSMICRAYS from the CRUTIL package. 
Significant cosmic rays persisted in regions with strong emission
features. Any spurious feature that made it to the final 1-D extraction could
later be identified and removed by interpolation.

Each science exposure together with a flat and an arc frame were then
sent to a generic spectroscopy reduction pipeline written by D. Kelson
\citep{Kelson2003}.{\footnote{D. Kelson, The Observatories, Carnegie
Institute of Washington, 813 Santa Barbara Street Pasedena, California
91101 USA.}

This software divided the science images containing the 2D spectra by
a master flat.  The 2D spectra were then rectified to correct for any
distortion along the spatial and dispersion axes.  Using a He-Ar arc
frame the programme found the dispersion relation and wavelength
calibrated the 2D spectra.  The calibrated 2-D science frame output
from the Kelson software was then sent to the APALL programme in IRAF
which performed a variance weighted extraction.

The output of APALL for each slit were 4 1-dimensional arrays: The
weighted object spectrum, the unweighted object spectrum, the
subtracted sky background and a variance spectrum, all as a function
of wavelength.

Spectra were then flux calibrated using the IRAF tasks STANDARD and
SENSFUNC.  The standard star data was taken by ESO on the night of
the observations.

\subsubsection{Identification of emission lines}
The plotting package SPLOT was used to identify emission lines by
eye in the 1-D spectra.

Upon successful identification of any spectral features the galaxy was
sent to the NOAO programme RVIDLINES. Each emission line was fitted with a
Gaussian profile that was centred on the emission peak.  The FWHM of
the profile was set to a default 4 pixels (9.72 \AA), with a 5 pixel
minimum gap between 2 features. 

A variance weighted mean redshift was then computed using all of the
features identified. The heliocentric correction was made to the
redshift using the header information. An RMS error in the redshift was
also calculated using the redshift of each feature z$_{\rm{feature}}$
and $\rm{\bar{z}}$ in equation \eqref{eq:RMSerrorZ}.
\begin{equation}
\label{eq:RMSerrorZ}
\rm{\sigma_z}=\sqrt{\frac{\sum_{i=1}^{n}(z_{feature}-\bar{z})^2}{n-1}}
\end{equation}

Figure \ref{fig:HEM1B36} shows a typical spectrum of a galaxy
where only [O II] emission was used to determine the redshift.
H$\beta$ and [O III] lines were lost in the noise. The `4000 Angstrom break'
in the SED helped 
to identify this line as [O II] emission and not H$\alpha$. 

\begin{figure} 
\begin{center}%
\includegraphics[scale=0.50]{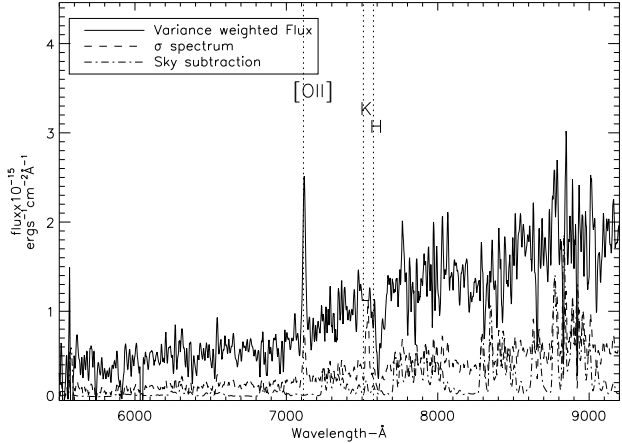}
\caption{A variance weighted combined spectrum of slit 36 from
``bright'' mask 1 of sightline HE 1122-1648. The redshift of this
emission line dominated galaxy was determined using the [O II]
emission line only. The H and K absorption lines 
were located near a cosmic ray which has been interpolated over
at $\sim7500$ \ang. The feature at $\sim7600$ \ang\
is atmospheric O$_{2}$ absorption. For display purposes the
spectrum has also been box-car smoothed with a window of 5 pixels.
The residuals at $\gtrsim9000
\rm{\AA}$ are a product of poor sky-subtraction
and so the H$\beta$ and [O III] doublet could not
be identified. The increase in flux above $7600$ \ang\ meant 
the emission line could be identified as [O II] rather
than H$\alpha$ because of the `4000 Angstrom break'. The
$\sigma$-spectrum (dashed) and sky value (dot-dashes ) that was
subtracted are also shown.  The value of the sky spectrum has been
divided by 100.}
\label{fig:HEM1B36}
\end{center}
\end{figure}

Using this method the redshifts of 180 emission-line dominated
galaxies were found.

\subsubsection{Identification of absorption lines}
Poor signal-to-noise (S/N) meant identifying the redshifts of
absorption-dominated galaxies was not possible by inspection.  Instead
the IRAF programme XCSAO was used \citep{Kurtz1998}. Absorption
features were identified by cross-correlating with a template
absorption spectrum. These absorption lines were then used in
RVIDLINES to measure redshifts for the 13 absorption-line dominated
galaxies that were included in the final table.

\subsubsection{Redshift Confidence of the Galaxies}
\begin{table}
\caption{Confidence levels and criteria needed for the galactic redshifts}
\begin{tabular}{clc} \hline
\hline
Confidence & Comment & No. of \\
Level & & Galaxies \\ 
\hline
1 & 2 or more strong lines & 54 \\
2 & 2 or more moderate lines & 26 \\
3 & 2 or more weak lines & 18 \\
4 & Secure single line [O II] redshift & 61 \\
5 & 2 or more weak lines from poor spectra & 6 \\
6 & Ambiguous single line [O II] redshift & 28 \\
\hline
\label{tab:emabslines}
\end{tabular}
\end{table}

Each of the 193 galaxies that made up the final table were then
assigned a redshift confidence level based on the number and strength
of any features. The criteria for each level are shown in Table \ref{tab:emabslines}.

Those galaxies with a ranking of 1 were most often AGN. The galaxies
dominated by absorption lines were assigned a ranking of 2 or 3
depending on how visible the lines were after being identified in
XCSAO. 61 confident redshifts were found, even though only [O II]
emission was visible because of features such as the `4000 Angstrom break'.

The 28 [O II] emission lines for which there was no alternative feature
could have been H$\alpha$. This meant there were a total of 34 spectra
(those ranked 5 or 6) for which there were ambiguous 
results. Figure \ref{fig:rankz} shows the redshift distribution
of the 193 galaxies binned according to whether the redshift was
judged to be reliable.

\begin{figure} 
\begin{center}%
\includegraphics[scale=0.50]{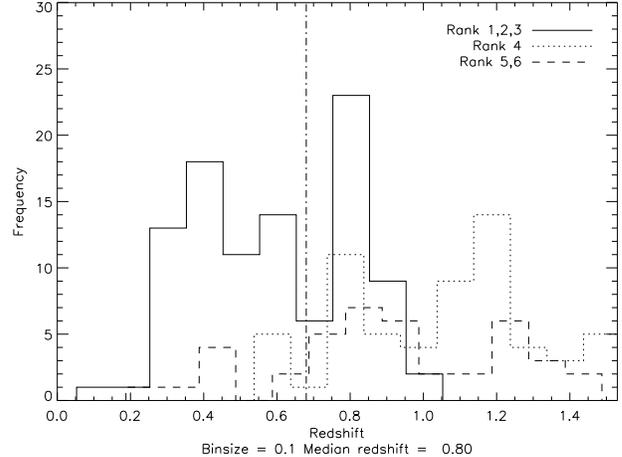}
\caption[The Redshift rank distribution]{The redshift distribution of
the 193 galaxies is shown in this figure, binned as a function of the
redshift confidence. This plot shows that most of the galaxies in the
galaxy sample used in the correlation were obtained from single line 
[O II] emission and had a
redshift $\rm{z}\gt0.8$. The dashed-dotted line at $\rm{z}=0.68$ shows
the minimum redshift at which galaxies could correlate with \lya\
absorbers found in STIS E230M spectra.}
\label{fig:rankz}
\end{center}
\end{figure}

Those galaxies with an insecure redshift were flagged when passed to the
code and removed from the calculation.
The 61 galaxies only showing confirmed [O II] emission were also flagged
in order that they too could be removed in case they significantly 
changed the result of the correlation.

Being restricted to effectively a single night's observation meant we
did not achieve a large completeness in the galaxy survey.  Figure
\ref{fig:completesep} shows the percentage of galaxies with a
determined redshift of the 1062 and 1122 objects found using
SExtractor. They have been plotted as a function of their separation
from the quasar sightline. This incompleteness is accounted
for in the calculation of the correlation function.

\begin{figure} 
\begin{center}%
\includegraphics[scale=0.50]{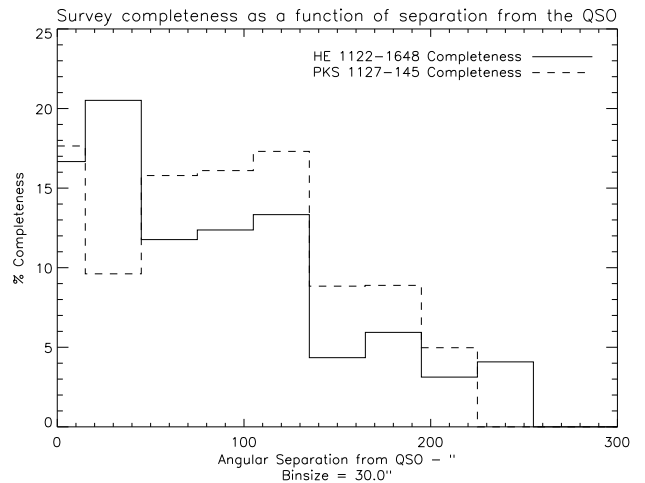}
\caption[Angular completeness distribution]{Of the 1062 and 1122 candidate galaxies detected
in the fields of quasars HE 1122-1648 and PKS 1127-145 this plot shows the percentage
of those with a determined redshift. The completeness is calculated for the apparent
magnitude ranges $21.5\le\rm{m}_{R}\le23.0$ and $21.5\le\rm{m}_{R}\le24.5$.}
\label{fig:completesep}
\end{center}
\end{figure}

\subsection{The Luminosity Distance and K-Correction}
\label{sec:kcorrect}
Using the selected $\Lambda$CDM cosmology stated in Section 1 the
luminosity distance to each galaxy ($\rm{D_{L}}$) was found and the
absolute magnitude $\rm{M}_B$ of each galaxy was computed.
\begin{equation}
\rm{M}_{\it{B}} = \rm{m}_{\it{R}} -5\log_{10}\left(\frac{D_{L}(z)}{10\;\rm pc}\right)+ K_{correct}
\end{equation}

m$_{R}$ took the value of MAG AUTO that was computed when
SExtractor was run on the pre-imaging. As the pre-imaging was
only exposed through the red Bessel filter a theoretical K-correction
had to be computed using the response curves of two filters and the
spectral energy distributions (SEDs) of template galaxies.  Details of
the code used for the colour correction and the template SEDs are
described in \citet{Bruzual2003}.

To compute the \br K-correction for each spectral type the theoretical
flux that would be transmitted through the rest-frame $B$ filter was
computed. Each SED was then transformed to a higher redshift in steps
of $\Delta\rm{z} = 0.005$ and the flux through the red filter was
integrated.  The output from the code was a table, with a value for
the K-correction listed as a function of redshift and spectral
type. Galactic evolution was not taken into account when transforming
the SEDs. With no clear indication from the galaxy spectra as to the
spectral type of each galaxy the \br colours for all 5 templates were
considered. Figure \ref{fig:absmagBz} shows the mean absolute magnitude
of all 193 galaxies based on the 5 SED templates. From Figure \ref{fig:absmagBz}
it is apparent that we were calculating the cross-correlation and galaxy auto-correlation
using galaxies that had $-17\gtrsim\rm{M}_B\gtrsim-21$. We did not have
a sufficient number of galaxies to investigate the dependence 
of the correlation on the galaxy magnitudes.

\begin{figure} 
\begin{center}%
\includegraphics[scale=0.50]{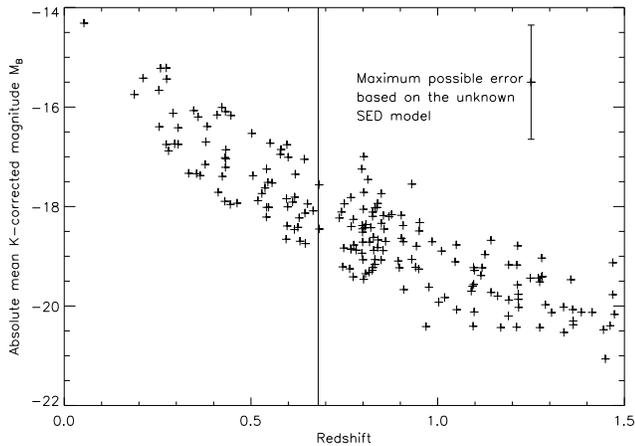}
\caption[The absolute magnitude of the galaxies]{The mean absolute magnitude M$_{B}$ of
the 193 galaxies based on the 5 \citet{Bruzual2003} SED templates. The error bar 
is equal to the largest range in the possible values of M$_{B}$ from all of the galaxies.
The vertical line shows the cut-off at $\rm{z}=0.68$ above which the galaxies were at an
appropriate redshift to correlate with the \lya\ absorbers and for use in the 
auto-correlation.}
\label{fig:absmagBz}
\end{center}
\end{figure}

Altogether the spectroscopic redshifts of 193 galaxies were found and
were added to a master catalogue. This included 37 in the field of
HE 1122-1648 that were at a redshift of $\rm{z}\ge0.68$ and had a redshift
confidence $\leq4$, and so could
be correlated with the \lya\ absorbers. The number of galaxies that could be used 
in the calculation that were found in the
field of PKS 1127-145 was 58.

\section{Quasar UV spectroscopy}
The \lya\ absorbers used in the cross-correlation function were
detected in the absorption spectra of the background quasars HE
1122-1648 and PKS 1127-145. These
quasars were selected because high resolution (R$=30000$), STIS
echelle spectroscopy obtained using the E230M grating exists within
the HST archive.  These archived spectra cover the wavelength range
2050-3111 \ang, the range of wavelengths needed to locate \lya\
lines at a redshift of $\rm{z}\sim1$.  With such a high dispersion, it
was hoped that low column density (\logcoldens$\sim13$) HI absorption
lines could be resolved and Voigt profiles fitted in order that their
column density and b-parameter could be determined.

The absorber sample in this paper has a higher resolution than the
absorber data set of \citet{M_J_2006}. They used HST FOS data and 
a fitted b-parameter that was fixed at $30$ \kms.
The STIS E230M instrumental broadening was 10 \kms\ in the spectra we used,
compared to 230 \kms\ for FOS spectra. Therefore in principle we
are sensitive to lower column density lines than \citet{M_J_2006}, and we can
directly measure the b-parameters for absorbers, rather than assuming
a value of 30 \kms.

The only \lya\ absorbers that were used in the correlation function
were detected using the Space-Telescope-Imaging-Spectrograph (STIS) in
echelle mode with the E230M grating. In addition to these, to help
identify lines and to compile a more complete line-list, spectra were
analysed that were taken by STIS in long-slit mode, the
low-resolution Faint Object Spectrograph (FOS) and the ESO UVES
(Ultraviolet and Visual Echelle Spectrograph) instrument.

Table 4 lists the data products used in the
investigation that were collected from the STScI (Space Telescope
Science Institute) and ESO (European Southern Observatory) archives
\citep{fosHandbook, STISHandbook, UVEShandbook}.

\begin{table}
\label{tab:archivedata}
\caption{Spectra from the ESO and HST data archives}
\begin{tabular}{cccc} \hline
\hline
Instrument + & $\lambda$ & Exp & $\frac{\lambda}{\Delta\lambda}$ \\
grating & (\ang) & time (s) &       \\
\hline
& HE 1122-1648&& \\
\hline
 STIS  E230M$^{\rm{a}}$ & 2270-3111 & 4x2877 & 30000 \\
 STIS  E230M$^{\rm{a}}$ & 1574-2382 & 6x2712 & 30000 \\
 STIS  G430L$^{\rm{a}}$ & 2900-5700 & 2x1000 & 1040  \\
 FOS   G190H$^{\rm{b}}$ & 1571-2311 & 3x1873 & 1300  \\
 UVES  Blue/Red$^{\rm{c}}$ & 3200-5765 & 15x5400 & 45000 \\
\hline 
& PKS 1127-145&& \\
\hline
 STIS  E230M$^{\rm{d}}$ & 2270-3111 & 14x5798 & 30000 \\
 FOS   G190H$^{\rm{e}}$ & 1571-2311 & 4x2160  & 1300  \\
 FOS   G160L$^{\rm{f}}$ &  901-2508 & 1x1450  & 250   \\
 UVES  Blue/Red$^{\rm{g}}$ & 3200-6809 & 10X3060+4x4800 &  45000 \\ \hline
\end{tabular} \\
a) PI Baldwin HST proposal 9885
b) PI Reimers HST proposal 5950
c) \citet{UVESHE1122,Kim2002}
d) PI Bechtold HST proposal 9173
e) \citet{Deharveng1995}
f) \citet{RaoG160L}
g) \citet{UVESPKS1127}
\end{table}

\subsection{Reduction of both long-slit and echelle spectra}
The preliminary calibration and reduction of data for each instrument
was performed by STScI or ESO `on-the-fly' by an instrument specific
pipeline. The steps performed in the initial reduction of raw data
depended on the detector.

\subsubsection{FOS calibration}
The details for the FOS pipeline CALFOS are from \citet{fosHandbook}
and they differed significantly from both the STIS and UVES pipelines
as the data was already 1-dimensional.  The pipeline began by
computing the statistical error in the raw data, this was then carried
forward through each reduction step and used in the variance spectrum.
The total photon count was divided by the exposure time to give the
count rate.  Next the background sources of noise were
subtracted. This included the dark current generated by the instrument
and scattered light from the optics.  The variations in sensitivity
with $\lambda$ from diode-to-diode and the illumination correction
were then compensated for by dividing each pixel by the flat-field
response.  After each observation FOS would perform an arc exposure
using the internal Pt-Cr-Ne lamp in order to find the wavelength
zero-point. Vacuum wavelengths were used and the pipeline, using
the date of observation, performed a heliocentric velocity correction.
Finally using the data from a spectrophotometric star the
pipeline flux calibrated the object spectrum converting the count rate
to $\rm{ergs\;s^{-1}\;cm^{-2}\;\rm{\AA}^{-1}}$, thereby correcting for
the telescope and spectrograph transmission and the detector
efficiency.

\subsubsection{STIS calibration}
The initial stage of the pipeline CALSTIS varied depending on whether
the data was taken using the MAMA or CCD detector.  A detailed
description of both pipelines is covered in \citet{STISHandbook}.
Similar to the FOS pipeline the first process was to calculate a
statistical error, this error was then propagated through the
reduction pipeline.  The dark current was then subtracted and the 2D
spectra were divided by a flat field exposure created by a lamp within
the spectrograph.  The STIS CCD data was reduced in the same manner as
generic CCD optical data.  For both UV and optical data the STIS
pipeline then used an arc exposure to calculate the wavelength
dispersion solution with the necessary heliocentric velocity
correction.  Using a spectro-photometric star the spectra were also
flux calibrated to give counts in
$\rm{ergs\;s^{-1}\;cm^{-2}\;\rm{\AA}^{-1}}$.  The 2D long-slit and
echelle data were then binned along the spatial axis to produce 1D
spectra.

\subsubsection{UVES Calibration}
The pipeline steps for the UVES raw data followed the same procedure
as the STIS optical data save for two main differences. As UVES is a
ground based instrument the sky needed to be subtracted when
extracting the 1D spectra.

When we analysed the absorption lines of the UVES spectra we used
vacuum wavelengths so that they would correspond to the absorption
lines of the STIS data.  Therefore the dispersion solution needed to
undergo an air-to-vacuum wavelength correction before
transforming to the heliocentric reference frame.

\subsubsection{Combining multiple exposures}
We performed the final steps in reducing the 1D spectra.  We combined
exposures using variance weighting and fitting a QSO continuum.  A
different method was used depending on whether the spectrum was
long-slit data or from echelle spectroscopy.

The nature of digicon, MAMA and CCD detectors means that, prior to any
combining of exposures, each of the spectra had a wavelength relation
that was not the same or constant as $\Delta\lambda\propto\lambda$.
So the first step before combining multiple exposures was to inspect
their dispersion and then re-sample each of the spectra to a common
wavelength solution.

\subsubsection{Fitting the QSO continuum}
When analysing absorption lines the QSO flux needed to be divided by
the QSO continuum.

The continua for the long-slit spectra were fit using the IRAF programme
SPLOT.  The first step when estimating a continuum was to crop the
spectrum by $\sim100\;\rm{\AA}$ at both the red and blue ends. This
was to remove data with a S/N per pixel $\leq 2$. Any emission
features or large absorption profiles were also clipped. In particular
this included the \lya\ and \lyb\ emission lines if they were
present. The continuum of a QSO spectrum was fitted by sampling the
spectrum at points devoid of absorption, usually in regions redwards
of the \lya\ emission, this was then extrapolated to lower
wavelengths. A chebyshev function of order 2 was fit to the data using
SPLOT. The final spectra were then saved in a FITS format for use in
RDGEN, a programme that would normalise the spectra by dividing through
by the continuum and detect absorption features. RDGEN is part of the
VPFIT\footnote{\copyright 2007 R.F.Carswell, J.K. Webb, M.J. Irwin,
A.J. Cooke, Institute of Astronomy, Madingley Rd, Cambridge, CB3
0HA,Uk.}  package that was used when modelling Voigt profiles.

\subsection{Post calibration STIS E230M and UVES Echelle spectroscopy reduction}
The echelle data was composed of many
exposures, these not only needed to be re-sampled and combined, but
also the separate overlapping orders needed to be joined end-to-end
while simultaneously avoiding aliasing or losing flux when
re-sampling.  For this reason a C program called UVES popler
(version 0.17) \citep{MurphyUVES_POPLER} was used. This was devised
and written by M. T. Murphy \footnote{ M. Murphy, Centre for
Astrophysics \& Supercomputing, Swinburne University of Technology,
Mail 39, PO Box 218, Hawthorn, Victoria 3122, Australia} with the
intention of reducing post-pipeline UVES echelle data.  The input for
UVES popler were all the optimally extracted echelle orders from the
STScI CALSTIS and UVES pipeline and the associated variance spectra.
Combining these orders, resampling the data, setting a common
dispersion solution and normalising by the QSO continuum 
was then an automated process.

Parameters were passed to UVES popler that gave the function and order
to fit for the continuum. A chebyshev function of order 2 was selected
above and below the \lya\ emission line.  High column density
absorption lines were ignored by UVES popler by rejecting any data
that was in the lower 50\% of the flux values below the \lya\ emission
line. Flux values that made up the lower 30\% of the data above \lya\
were also ignored.  A higher tolerance was permitted at these
wavelengths as there were only a few metal absorption lines and no
forest absorption.

The level to set the continuum was then iteratively improved by
neglecting data that remained 1.2 $\sigma$ below a provisional
continuum line and any emission/noise features that existed $3 \sigma$
above.  We visually inspected the final continuum, adjusting it by
hand where necessary. Errors in the  continuum level may introduce
small errors in the inferred column  densities of lines for regions
of our spectrum with low S/N ($\leq3$), but we do not expect these
errors to affect our measurement of the line redshifts.

Both HE 1122-1648 and PKS 1127-145 were fit with the same parameters
for the STIS and UVES spectra. 

\subsection{Producing a Quasar Absorption line-list}
Locating the \lya\ absorbers that would be used in the
cross-correlation function at a redshift $\rm{z}\sim1$ required the
STIS E230M data. The other spectra were used for completeness, for the
identification of metals and to corroborate the identity of STIS \lya\
absorption lines.

The \lya\ and \lyb\ emission lines of quasar PKS 1127-145 are located
at $2658.67\;\rm{\AA}$ and $2243.25\; \rm{\AA}$. Therefore only the
\lya\ series lines were within the wavelength range of the E230M data.
\lya, \lyb\ and Ly-$\gamma$ emission lines for quasar HE 1122-1648 were
at $4133.28\;\rm{\AA}$, $3487.46\;\rm{\AA}$ and $3229.13\;\rm{\AA}$
respectively. Hence many Lyman series absorptions were located in the
UVES and STIS spectra. Locating these lines first was essential so
that higher order Lyman lines could be identified and removed. Any
remaining lines could then only be \lya\ absorptions (or metals).

We used RDGEN to locate absorption lines. Using the velocity FWHM of
the instruments; that were 10.0 \kms, 6.7 \kms\ and 230 \kms\ for the
STIS E230M, UVES and FOS G190H gratings respectively.
The output from this was a table that included the wavelength, the observed
equivalent width (EW), the error associated with these parameters and the
significance of the absorption line.

The significance level of each line was defined as, given error in
the EW $\sigma_{\rm{EW}}$.
\begin{equation}\label{eq:linesignif}
\rm{SL}=\frac{\rm{EW}}{\sigma_{\rm{EW}}}
\end{equation}
At this stage all the lines within the STIS spectra where RDGEN had
estimated an EW significance $\sigma\geq 3$ were passed forward to the
line-fitting programme VPFIT.  No contribution to the error in the EW
caused by the error in the level of the continuum was included at this
point.

Absorption lines were fitted with Voigt profiles using VPFIT (version
9.3).  Using the instrumental velocity FWHM VPFIT would attempt to model the
absorption line. This was achieved by a $\chi^2$ minimisation when
comparing the data and fitted Voigt profiles.  The parameters of
the lines were then returned together with $1\sigma$ errors.  The
input to the programme was a table of the absorbers detected using RDGEN
that included the ion species and an approximate initial guess for the
redshift, column density and b-parameter. The atomic data required was
taken from \citet{Morton_2003}.

The following steps were made when identifying and then modelling the
lines. First the galactic lines such as \MgII\ and
\FeII\ were identified and flagged as any redshifted line that
was blended with these had to be ignored. 

The second step was to identify redshifted metal lines above the \lya\
emission line.  No redshifted metal lines were
detected at any wavelength in the STIS spectra. Those detected in UVES
were associated with a damped (\logcoldens$\geq20$) system.  The HE
1122-1648 FOS G190H data exhibited a damped system at $\rm{z}=0.681$
(\logcoldens$=20.5\pm0.03$).  \CIV, \MgII\ and \FeII\
absorption lines located in the UVES spectrum are associated with this.

The other method employed to identify metal lines such as
\CIV\ and \MgII\ was by their doublet wavelength separation.

Metal lines were found within the Lyman-series forest, in
particular the doublets \CIV\ ($\lambda\lambda1548.20,1550.77$)
and \OVI\ ($\lambda\lambda1031.93,1037.62$).  As this was done by eye
only two ways were known to discriminate between metals and possible
\lya\ lines. First if both lines could be spotted there was the
doublet spacing. In addition, for metal lines the b-parameter is usually
$\leq10$ \kms\ while for \lya\ b it is seldom $\leq15$ \kms\ and
cannot be $<10$ \kms\ \citep{Janknecht2006}.  So an inspection of the
FWHM gave a good indication as to the nature of the species.

With the metal absorption lines identified it could now be assumed
that those left were Lyman lines. First of all, any remaining
unidentified lines within the \lya\ forest were assumed to be
HI. RDGEN was then used to plot the output of VPFIT and the location
of the \lyb\ lines were highlighted.  VPFIT was then rerun if there
was a positive match so that both lines could be used to constrain the
errors.  Any line remaining unidentified within the \lyb\ forest could
only then be \lya.  This process was repeated towards shorter
wavelengths until the blue end of the STIS echelle spectrum was
reached. In the case of PKS 1127-145 only the \lya\ forest was
visible. The damped system at $\rm{z}=0.681$ in
the HE 1122-1648 spectrum prevented any further lines being discovered
at shorter wavelengths.

\subsubsection{The minimum detectable column density}
An effective way to decide the minimum detectable EW for the \lya\
lines was to invert the spectrum around the continuum, and using the b
value of the smallest real \lya\ absorber we attempted to fit an
`absorption line' to the inverted spectrum. A line that we knew was a
sample of the noise. When the column density and $1\sigma$ error to
the largest noise profile became indistinguishable from that of the
smallest known real absorber this gave the minimum reliable value. For
the PKS1127-145 and HE 1122-1648 STIS spectra the minimum detectable
column densities were \logcoldens$= 13.2$ and \logcoldens$= 13.5$
respectively.  For the UVES HE 1122-1648 spectrum where the S/N was
$\sim120\;\rm{pixel^{-1}}$ at the central wavelength \citep{1122s_N}
this minimum decreased to \logcoldens$= 11.7$.

VPFIT would sometimes attempt to fit these weak lines with unphysical
b-parameters. As the Doppler parameter is
$\propto\sqrt{\frac{2\rm{kT}}{\rm m}}$ any \lya\ line that was fitted
with $\rm{b}\leq 10$ \kms\ was dropped from the list, where we have
adopted the same cut that was used in \citet{Janknecht2006}.

\subsection{Extracting lines from data with very poor signal-to-noise}
\subsubsection{Fixing the variables}
We set VPFIT to terminate if the $1\sigma$ error in the column density
was greater than $\Delta$\logcoldens$=0.5$.  Whenever this point was
reached we would fix the b-parameter to the median b value of those
lines that had been measured. This was 26.5 \kms.  This was done for
all of the \lya\ lines below $2300\;\rm{\AA}$ regardless of
$\Delta$log(N$_{\rm HI}$) in the HE 1122-1648 STIS spectrum where the
S/N was very poor.

\subsubsection{The blue HE 1122-1648 STIS E230M spectrum} 
\label{sec:noisydata}
The S/N of this spectrum was below 2 for most
wavelengths.  The wavelength range was meant to cover
$1574\rightarrow2382\;\rm{\AA}$ but the data became unusable below the
damped system at $2050$ \ang. 

In an effort to recover the absorption features in this low S/N data,
we summed the flux over a narrow bandpass of 24 pixels and then
plotted this as a function of wavelength. The wavelength of the
absorption lines that were lost unbinned data could then be
determined. Knowing where these lines were meant that Voigt profiles
of a fixed b-parameter could then be fit.

The spectrum was cropped to remove those wavelengths below the DLA
\lya\ line.  The flux absorbed ($\rm{d_{i}}$) per pixel i was
then summed over a window of size M+1 using the following calculation.
C$_{\rm{i}}$ is the continuum level normalised to 1 for all i and
F$_{\rm{i}}$ is the flux.

\begin{equation}
\label{eq:fluxdeficit}
\rm{d}_{i}= \sum_{j=-\frac{\rm{M}}{2}}^{\frac{\rm{M}}{2}-1}\left(\rm{C}_{i+j}-\rm{F}_{i+j}\right)
            \times(\lambda_{i+j+1}-\lambda_{i+j})
\end{equation}

The variance of this result was also computed where the values were
summed in quadrature.
\begin{equation}
\sigma_{\rm{i}}= \sqrt{\sum_{\rm{j}=-\frac{\rm{M}}{2}}^{\frac{\rm{M}}{2}-1}
           \left[\sigma_{i+j}\left(\lambda_{\rm{i+j+1}}-\lambda_{\rm{i+j}}\right)\right]^2}
\end{equation}

The summed absorbed flux as a function of wavelength is plotted in
Figure \ref{fig:fluxdef}a. 

\begin{figure*}
\begin{center}
\includegraphics[angle=90.,scale=0.60]{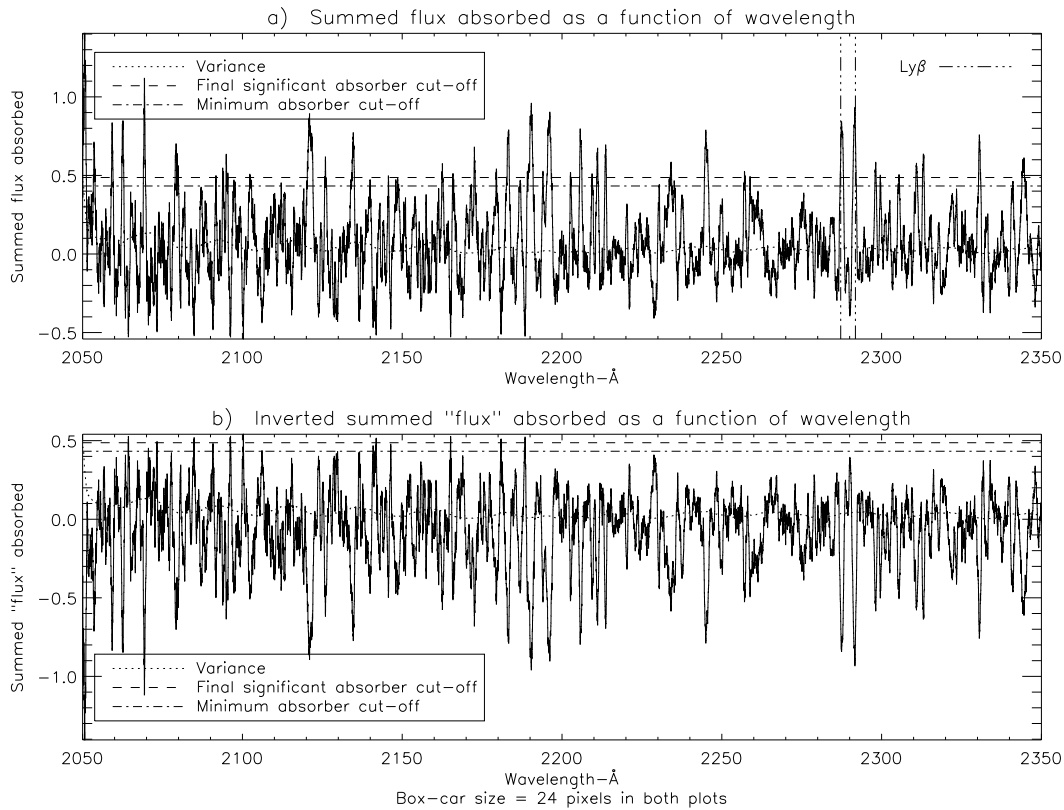}
\caption
{(a) In order to locate \lya\ absorbers in the blue STIS E230M
spectrum of HE 1122-1648 the absorbed flux was summed over a window of
24 pixels using equation \eqref{eq:fluxdeficit}.  This way any
absorbers were enhanced and could be modelled using VPFIT.  (b) The
`significant absorber' cut-off lines were set by inverting about the
continuum the spectrum and sampling the noise. To recover the expected
line-density at this redshift would require a cut in the flux deficit
of 0.43, however many noise features would then be included
(dashed-dot line).  Thus a more conservative cut of 0.49 was chosen.  After
a final EW significance level cut $\rm{SL}\ge3$ 31 \lya\ lines were
identified.}
\label{fig:fluxdef}
\end{center}
\end{figure*}

Two identifications were made in this spectrum. These
were two \lyb\ lines associated with the systems at $\rm{z=1.229}$ and
$\rm{z=1.234}$, these can be seen in Figure \ref{fig:fluxdef}a at
$\sim 2290\;\rm{\AA}$. The smallest profile had a width of 24 pixels,
so assuming this profile which had an EW significance $\sigma= 8.7$, was
typical of the minimum width we could hope to identify, this was
chosen to be the window size M.

The next variable to find was the minimum detectable EW for a line.
For this we adapted the earlier method of inverting the spectrum to
sample the noise. Equation \eqref{eq:fluxdeficit} was applied to the
inverted spectrum using the same window size, this is plotted in
Figure \ref{fig:fluxdef}b.  The peak inverted summed `flux' had a
value of 0.54. Using this cut on the real summed spectrum meant only
28 absorbers would be included. So by using the peak noise value there
was a danger that many real \lya\ absorbers would be lost.

The red HE 1122-1648 STIS E230M spectrum had a \lya\ line density of
$\frac{d\rm{N}}{d\rm{z}}=198$ and all results indicate that the line
density distribution is almost flat at redshifts $\rm{z}\leq1$
\citep{Kirkman2007}, so a similar value is expected. The conservative
cut that left 28 absorbers implied a line density of
$\frac{d\rm{N}}{d\rm{z}}=165$. This cut-off needed to be reduced by
$20\%$ to a minimum flux-deficit of 0.43 before a similar line density
could be observed. Figure
\ref{fig:fluxdef}b shows such a minimum cut-off would allow a lot of
false identifications from smoothed noise.

Therefore the cut was varied between 0.43 and 0.54 in order to see to
what proportion of lines would be excluded when compared to the density of
\lya\ lines that are expected. Eventually the cut-off in Figure
\ref{fig:fluxdef}b was set to exclude all but the highest noise peaks
at 0.49. In doing this 46 absorbers were identified.  A programme was
then written that measured the wavelength of these peaks. Those
that were higher order HI lines or had an EW significance 
less than 3 were removed. This left 31 \lya\ absorbers that could be
measured using VPFIT.

The original flux and variance spectra spectra were then smoothed
using a box-car window of 11 pixels and this was used in the VPFIT
programme.  VPFIT was now successful in modelling the Voigt profiles
with $1\sigma$ errors in the column density $\Delta$\logcoldens$<0.5$
and these absorbers were added to the final line-list. All of the
lines found using this method had a b-parameter fixed to the median
value found for those lines that had been fitted successfully at
higher wavelengths of $26.5$ \kms.

\subsection{Variance Weighted Equivalent Width Significance and the Continuum error}
All \lya\ lines with an equivalent width significance $\rm{SL}\geq3$
were included in the cross-correlation function, this is the level
adopted in \citet{KPew_measure}. However we had to fix the b-parameter
for 42 out of the 135 \lya\ lines in the STIS spectra that had an
equivalent width significance of 3, and assume the continuum was
correct. Therefore the EW and significance of the lines were measured
again. This time a Gaussian weighting was used where changes in the
significance of a line because of errors in the continuum were also
considered.  The formulae used to calculate the EW and significance
have been adapted from \citet{KPew_measure}.

Consider an absorption line that covers $\rm{J}$ pixels and is 
centred on pixel i with the continuum C set to 1. The EW of the 
profile is equal to: 
\begin{equation}
\label{eq:KPew}
\rm{EW} = \frac{\Delta\lambda_{i}\sum_{\rm{j=1}}^{\rm{j=J}}\eta_{j}
               \left(C_{i+j-J/2}-F_{i+j-J/2}\right)}
               {\sum_{\rm{j=1}}^{\rm{j=J}}\eta^2_j}
\end{equation}

$\Delta\lambda_{\rm{i}}$ was assumed to be constant across the
absorption, and $\eta_{\rm{j}}$ is the Gaussian weighting coefficient
where $\rm{\sum_{j=1}^{j=J}\eta_j=1}$.  The FWHM
of the Gaussian profile for each absorber was based on its
b-parameter.

The amount that the variance spectrum ($\sigma^{2}$) contributed
to the error in the EW was computed using:
\begin{equation}\label{eq:KPewVARerr}
\rm{\delta \left(EW_{\,var}\right)} = \frac{\Delta\lambda_{i}
                \sqrt{\sum_{\rm{j=1}}^{\rm{j=J}}\eta^2_{i}\sigma_{\rm{j}}^2}}
                {\sum_{\rm{j=1}}^{\rm{j=J}}\eta^2_j}
\end{equation}
The significance of each line was then, like equation
\eqref{eq:linesignif}, the ratio of equations \eqref{eq:KPew} and
\eqref{eq:KPewVARerr}.

Our approach when testing the effect of changes in the level of the continuum
was to estimate the significance levels when an error
$\rm{\delta\left(C_{\rm{j}}-F_{\rm{j}}\right)}$ of 10\% was added.
VPFIT was not rerun to find new values for the b-parameter
or the column density.
\begin{subequations}
\begin{align}
\left(\delta \rm{EW_{j\,cont}}\right)^2 &= \rm{\left(\frac{\partial EW_{j}}
                                          {\partial\left(C_j - F_j\right)}\right)^2
                                          \left(\delta\left(C_j - F_j \right)\right)^2}\\
\therefore\;  \left(\delta \rm{EW_{\,cont}}\right)^2 &= \rm{\sum_{\rm{j=1}}^{\rm{j=J}}
                                          \left(\frac{\Delta\lambda_i \eta_j}
                                          {\sum_{\rm{j=1}}^{\rm{j=J}}\eta_j^2}\right)^2
                                          \times0.01}\label{eq:conterr}
\end{align}                                          
\end{subequations}

Errors in the EW, that included contributions from both the continuum
and variance, were calculated using equations \eqref{eq:KPewVARerr}
and \eqref{eq:conterr} summed in quadrature.  Those absorbers that had
a new variance weighted equivalent width significance $\sigma\geq3$
were used in the cross-correlation.

It was found that even with a 10\% error margin in the continuum, the
Gaussian weighted significance of all the STIS absorption lines did
not change by more than $1\%$ and no additional lines were lost
because of the $3\sigma$ cut. This was because of the relatively high
minimum observable optical depth produced by a \lya\ column density
of \logcoldens$= 13.2$.

Figure \ref{fig:EWvsZ} shows the EW of the 135 lines detected
in the STIS spectra that had a significance level $\geq3$.

\begin{figure} 
\begin{center}%
\includegraphics[scale=0.50]{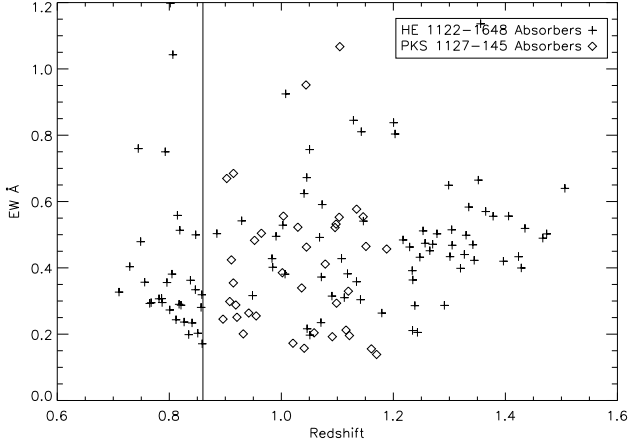}
\caption[EW vs Z for 135 absorbers] {The equivalent width of those
\lya\ absorbers with a significance $\sigma\ge3$ as a function of
redshift. The increasing minimum detectable EW of those absorbers from
opposite ends of the HE 1122-1648 STIS spectrum can be
observed. Those absorbers below the vertical line at $\rm{z}=0.86$ were detected
in the low S/N blue STIS spectrum.}
\label{fig:EWvsZ}
\end{center}
\end{figure}

One can observe that the peak in the redshift distribution of galaxies
in Figure \ref{fig:rankz} is at redshifts $\rm{z}\sim0.8$. Thus many
of the pairs in the correlation according to Figure \ref{fig:EWvsZ}
were made with those absorbers from the spectrum with poor S/N. These
are absorbers with a higher minimum EW limit hence a lower
completeness. Pairs from this region made up 72 of the 171 detected in
the cross-correlation plot in Figure \ref{fig:absgalran}a.  The effect
of this increase in the minimum EW means we may over-estimate the
cross-correlation. It is absorbers with a small EW that also have a
low column density, and we do not expect any correlation with
absorbers that have \logcoldens$\leq13.5$ \citep{ChenMulchaey2009}.

\subsection{The Final line-lists}
The final line-list for HE 1122-1648 contained 740 absorption lines
that were detected in the STIS E230M and UVES echelle data that
had an equivalent width significance greater than 3. This
included 586 \lya\ lines each with a b-parameter $\geq10$ \kms\ and
154 metals.  Of these \lya\ lines 97 came from the two STIS E230M
spectra.

Fewer \lya\ lines were detected in the STIS spectrum of PKS 1127-145
due to the lower quasar redshift. 38 \lya\ lines were found 
together with 23  metal lines in both the STIS and UVES spectra.  Therefore over
a total path length of $\Delta\rm{z}=1.08$ we detected 135 \lya\
absorbers that could be used in the correlation function.

The minimum column density for the \lya\ absorbers used in the
correlation function was \logcoldens$=13.2$.  The maximum column
density detected was \logcoldens$=17.4$, however only 11 absorbers out
of those correlated had a column density greater than
\logcoldens$=15$. The median column density was \logcoldens$=14.0$.
The complete line-lists for both HE 1122-1648 and PKS 1127-145 are
available in electronic format.

\section{The 2-point Cross-correlation function}
When investigating the relationship between gas and galaxies the
calculation for the 2D 2-point correlation was based on the estimator
devised by \citet{DavisPeebles83}.  The 2D 2-point correlation
function ($\xi_{\rm{AG}}\left(\sigma\rm{,}\pi\right)$) in equation
\eqref{eq:2d2pcfb} gives the measure of
association between the \lya\ absorbers and galaxies across both the
projected ($\sigma$) and line of sight ($\pi$) separations by
comparing the real distribution of pairs with random catalogues. 

\begin{equation}
\xi_{\rm{AG}}\left(\sigma\rm{,}\pi\right) = \rm{N_{QSO}}\frac{DD}{RR}-1 \label{eq:2d2pcfb}
\end{equation}

Along each line of sight every absorber was paired with every galaxy
and the impact parameter and line of sight separation in co-moving
coordinates was calculated for each pair. Each random absorber was also
paired with every random galaxy and the separation found between these pairs.  
DD and RR are the number of real-real and random-random pairs respectively in each
bin along the $\pi$ and $\sigma$ separations.

The normalisation constant N$_{\rm QSO}$ fixed the total number of
random pairs to the same value as the number of real pairs across all
$\sigma$ and $\pi$.  This way
$\xi_{\rm{AG}}\left(\sigma\rm{,}\pi\right)$ gave a value for the
fractional excess of pairs compared to a random distribution. Due to
the different number of real pairs that occur along each quasar
sightline a separate normalisation constant was calculated for each
field before the pairs were binned together.

To compare the values of $\xi_{\rm{AG}}$ against a known observable
the galaxy auto-correlation function $\xi_{\rm{GG}}$ was calculated
using the same techniques as the cross-correlation.

\subsection{Measuring separations}
\label{chap:pairing}
The line of sight co-moving distance ($\rm{D_C}$), to both galaxies
and absorbers that are at a redshift z, was calculated using equation
\eqref{eq:losdistance}.

\begin{equation}
\label{eq:losdistance}
\rm{D_C}=\frac{\rm{c}}{\rm{H_0}}\int\limits_{0}^{z}\frac{d\rm{z^\prime}}
         {\sqrt{\Omega_{\rm{M}}\left(1+z^\prime \right)^3+
         \Omega_\Lambda}}
\end{equation}
The line of sight separation, $\pi$, was then 
$\pi = \vert\rm{D_{C\;galaxy}}-\rm{D_{C\;absorber}}\vert$.

The angular separation between 2 points on an image that have been
projected from the celestial sphere is given by equation
\eqref{eq:angdistance}, where $\delta$ and $\alpha$ are the 
declination and R.A of the targets in radians.

\begin{equation}
\label{eq:angdistance}
\rm{D}_\angle=\arccos\left(\sin \delta_{\rm{abs}}\sin\delta_{\rm{gal}}+  
            \cos \delta_{\rm{abs}}\cos\delta_{\rm{gal}}\cos
             \left(\alpha_{\rm{gal}} - \alpha_{\rm{abs}}\right)\right)
\end{equation}

The RA and DEC of each absorber were the J2000 coordinates of the
quasar in question retrieved from the SIMBAD database.  Using the mean
line of sight distance towards both the absorber and galaxy, the
impact parameter was then calculated to be $\sigma =
\rm{D}_\angle\overline{\rm{D_C}}$.  As long as the $\pi$ separation
was $\leq 20$ \hMpcsev, each real galaxy-absorber pair was then written out
to a file for subsequent binning.

\subsection{Generating the random catalogues}
\subsubsection{The random galaxies}
When producing the random catalogues it was important that the random
galaxies have the same position on the sky as the data in order to
keep the mask design information. Therefore adopting the technique
used in W07, each real galaxy that lay in the redshift range of the
STIS E230M spectra of that particular line-of-sight was replaced with
200 fake galaxies in that position on the sky.  

In order not to bias the 2D correlation function it was also important
that the redshift distribution of the random galaxies be the same as
the real sample. To achieve this the redshift of each randomly
generated galaxy was a perturbation of
$\delta\rm{z}=\pm0.01\rightarrow0.05$ from the galaxy that spawned it.
This range was chosen so that no random pair
generated from the same real pair would appear within
the plotting space of  $\Delta\pi=20.0$, $\Delta\sigma=5.0$ \hMpcsev.
For example there are 37 galaxies in the field of HE 1122-1648 between
redshifts $\rm{z}=0.68\rightarrow1.51$, hence there were
$37\times200=7400$ random galaxies.

\subsubsection{The random absorbers}
The random absorbers were generated in the same fashion. Each real
absorber was replaced with 200 randoms, each with the same column
density, b-parameter and EW as the original. 

So that the random absorbers would have the same line-density as the
real sample, the redshift of each random absorber was also a
perturbation from the original. $\delta\rm{z}=\pm0.06\rightarrow0.1$.
The separation between all random absorbers and random galaxies was
then collected and likewise binned if $\Delta\pi\le20.0$ \hMpcsev.

Using 200 random targets per real absorber or galaxy was decided on a
trial basis.  This value was a compromise between limited computing
power and ensuring that the RR pair distributions were sufficiently
smooth, with little variation in the number of pairs along the line of
sight at separations of equal impact parameter.

\subsection{Binning for the 2D plot}
We selected pairs separated by less than or equal to 20 and 5
co-moving \hMpcsev\ along the $\pi$ and $\sigma$ directions
respectively and binned them on a grid. Each value in this array was
then divided by the equivalent bin in the RR array and 1
subtracted to give $\xi_{\rm{AG}}\left(\sigma\rm{,}\pi\right)$.  Our
choice of binning was $\Delta\sigma=1.0$ \hMpcsev, and $\Delta\pi=2.0$
\hMpcsev. This was coarser than the earlier studies of RW06 and W07,
who binned $\Delta\pi=\Delta\sigma=0.1\;\rm{h^{-1}_{100}}\; \rm{Mpc}$
and $\Delta\pi=2.0\; \Delta\sigma=0.4\;{h^{-1}_{70}} \rm{Mpc}$.  We
required this coarser sampling because of the small number of real
pairs.

\subsection{The error calculation}
In the previous studies of RW06 and W07, the uncertainty in $\xi$ was
calculated by jackknife resampling of all the sightlines.  The
correlation function was recalculated removing one quasar at a time
and an error determined for each bin.

Having only two quasars in this study an alternative method was chosen
where mock galaxy tables and linelists were created with the same
number of absorbers and galaxies and redshift distribution as the real
data.  The generation of mock redshifts was identical to that used to
generate the random catalogues described above.  These were then
treated as real data sets and run using the exact same code. 50 of
these mock linelists and galaxy tables were produced together with 50
estimates for the correlation function made by removing one of the
mock absorber and galaxy catalogues at a time.\footnote{To avoid
confusion `mock' catalogues are ones treated as real data in order to
estimate an error and all have an independent value for DD.  Random
catalogues, be they from the real or mock data, are used in the
denominator of the estimator.} A RMS error based on the sample
variance was then calculated using equation \eqref{eq:myerror}. With
$\rm{N=50}$ mock datasets,

\begin{equation}
\label{eq:myerror}
\sigma_{\xi\left(\sigma\rm{,}\pi\right)}
         =\sqrt{\frac{\sum\limits_{i}\left(\overline\xi-\xi_i\right)^2}{\rm{N}-1}}
\end{equation}

The Poisson error was found to overestimate the significance
of the correlation result (W07), and so was not calculated.

\section{Results}
\subsection{The Galaxy Impact Parameter Distribution}
\label{sec:impactdist}
The maximum possible impact parameter between \lya\ absorber and
galaxy pairs is a function of redshift and was constrained by the
$6.8\arcmin \times5.7 \arcmin$ field of view of the FORS2 detector.
Figure \ref{fig:velsepschem} shows a plot of the maximum possible
impact parameter.
\begin{figure}
\includegraphics[scale=0.50]{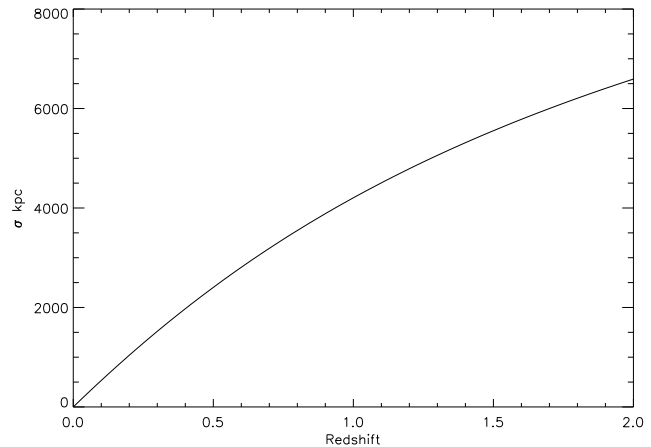}
\caption{The curve in this diagram shows the increase in the maximum
  possible observable impact parameter when using the $6.8\arcmin
  \times 5.7\arcmin$ detector on the FORS2 instrument.  This plot was
  derived using our stated cosmology and assuming that the galaxy was
  located in the corner of the detector, the angular separation
  $\delta\theta^2=3.4\arcmin^2+2.85\arcmin^2$ from the QSO at the
  centre of the chip.  The constraints of the size of the field and
  the maximum galaxy redshift meant we were able to correlate galaxies
  with absorbers out to a projected distance of $\sigma=5$ \hMpcsev.}
\label{fig:velsepschem}
\end{figure}
The solid line in Figure \ref{fig:velsep} is a histogram of the
projected separation between all 193 galaxies in the survey 
and the central quasar sightline.

The initial shape is caused by the increase with redshift in the
number of galaxies located within each concentric cylinder centred on
the quasar.

This distribution then peaked along a plateau at a projected separation
of $\sigma=800-2000$ \hkpcsev\ before rapidly declining above $\sigma=3000$
\hkpcsev.  This flattening of the distribution reflects the peaks in the
histogram of Figure \ref{fig:rankz} and the expected range in
impact parameter at these redshifts (see Figure
\ref{fig:velsepschem}).

Figure \ref{fig:rankz} shows most of the galaxies were detected at a
redshift $\rm{z}\sim0.8$.  Therefore the peak in Figure
\ref{fig:velsep} must occur below $\sigma\sim3000$ \hkpcsev. The
number of galaxies per bin had already started to decline by
$\sigma=2000$ \hkpcsev.  This was caused by a decrease in the radial
density of slits as the edge of the rectangular $6.8\arcmin \times
5.7\arcmin$ detector was reached.

The number of galaxies above $\sigma=4000$ \hkpcsev\ was low because the
size of the chip meant that a minimum redshift $\rm{z}\gtrsim1$ was
required to reach this impact parameter. 

\begin{figure}
\includegraphics[scale=0.50]{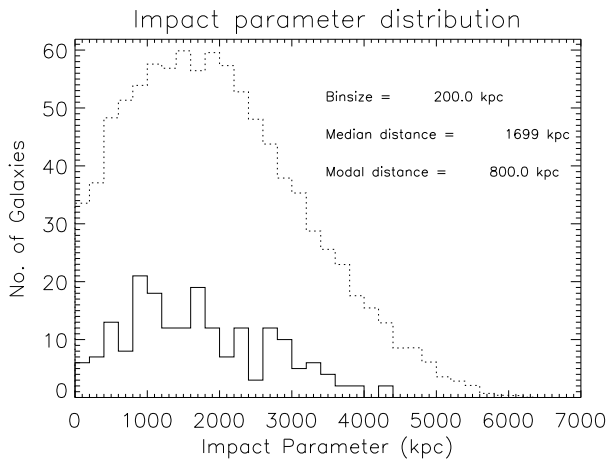}
\caption {The solid line histogram shows the number of galaxies
surveyed as a function of impact parameter binned by 200 \hkpcsev.  The
number of galaxies observed has an expected increase, peaking at
800-2000 \hkpcsev\ as the survey volume increases. The edge effects of the
chip and the drop off in the number of galaxies at high redshift then
cause a decrease in the number of galaxies beyond 3000 \hkpcsev.  The
dotted line represents the expected number of galaxies in a complete
magnitude limited survey ($21.5\leq R\leq24.5$) where the data is
from \citet{Cohen2000,Hogg97}. This plot shows that even though we
were able to sample the correct impact distribution for the galaxies
the survey was far from complete.}
\label{fig:velsep}
\end{figure}

The dashed histogram is a completeness test that shows the expected
separation had every galaxy within the same magnitude range
($21.5\leq$m$_{R}\leq24.5$) been targeted. Mock galaxies were placed
with a uniform spacing over the same $6.8\arcmin\times5.7\arcmin$
area. The required surface density for this magnitude cut came from
Figure 4 of \citet{Hogg97}.  Each mock galaxy was then assigned a
random redshift taken from the Hubble Deep Field North Survey redshift
distribution from Table 4 of \citet{Cohen2000}. A fiducial redshift cut-off
for the mock galaxies was chosen to be 1.8.  We then calculated the
impact parameter of each mock galaxy from an imaginary sight line
located at the centre of the CCD. This was repeated using 1000 mock
galaxy catalogues in order to calculate the expected impact parameter
distribution.

Figure \ref{fig:velsep} shows that our galaxy survey follows the same
general shape as the expected distribution. The mock catalogue also peaks in
the range $\sigma\sim1000-2000$ \hkpcsev\ before decreasing above $\sigma=2000$
\hkpcsev\ because of edge effects.  However the figure also shows that our
survey was far from complete.

\subsection{The \lya\ Absorber Redshift distribution}
Figure \ref{fig:linehist} shows a histogram of all the \lya\
absorption lines detected with an equivalent width (EW) significance
$\sigma\geq3$ from the STIS and UVES echelle spectra across both
lines-of-sight.

\begin{figure}
\includegraphics[scale=0.50]{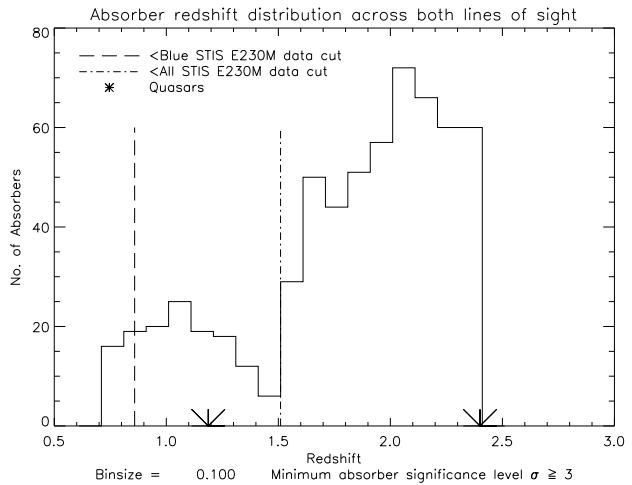}
\caption{This histogram binned by 0.1 in redshift shows the
distribution of all \lya\ absorbers with significance $\sigma \geq 3$
detected across both lines of sight and both the STIS and UVES
instruments. The dashed-dot line marks the redshift cut-off between
the two instruments and \lya\ absorbers above this were only found
along the sightline of HE 1122-1648.  \lya\ absorbers below the long
dashed line were found in the low wavelength HE 1122-1648 dataset, a
damped Lyman system prevented any more absorbers being detected with a
redshift below $\rm{z}=0.68$. No echelle STIS data below a redshift of
$\rm{z}=0.896$ existed for QSO PKS 1127-145.  The asterisks mark the
redshift of quasars PKS 1127-145 at z$=1.187$ and HE 1122-1648 at
z$=2.4$.}
\label{fig:linehist}
\end{figure}

All of the \lya\ absorbers below $\rm{z}=1.51$ were found in the STIS
data and are the 135 lines that were used in the cross-correlation
function. Those above are from the UVES spectrum and hence are all
from the HE 1122-1648 sightline.

Those below $\rm{z}=0.86$ were found in the blue STIS data set that
covered the wavelength range $2050-2382$ \ang and so are also only
from HE 1122-1648. The asterisks mark the redshifts of quasars PKS
1127-145 and HE 1122-1648 at $\rm{z}=1.187$ and $\rm{z}=2.4$
respectively.

Three factors affect the shape of this histogram.  The line-density
distribution only evolves slowly out to redshifts $\rm{z}\sim1.6$ and
then increases rapidly \citep{Janknecht2006}.  Our data agrees with
this result within the error margin. Our data is from only two
sightlines, one of which terminates at $\rm{z}=1.187$, but also the
deficit of lines at $\rm{z}\sim1.5$ is attributed to the decreasing
S/N and low sensitivity at the extreme red end of the STIS spectra of
HE 1122-16478 (see Figure \ref{fig:EWvsZ}).

The increase in the line density above $\rm{z}=1.5$ was also due
to the increased sensitivity and S/N of the UVES instrument.

\subsection{Combining the \lya\ Absorber and Galaxy datasets}
Figure \ref{fig:piplots} shows the location of both galaxies and absorbers
across each sight line.

The maximum impact parameter ($\sim5000\;$kpc) shows the distance out
to which the correlation function was calculated, and also explains
why so few pairs were found in the highest separation bin. Only those
galaxies with a redshift $\rm{z}\gtrsim1$ could achieve this
separation.

Pairs that were used for the correlation function had a maximum line
of sight separation of $20$ co-moving \hMpcsev. At a redshift of
$\rm{z}\sim1$ this equates to $\Delta\rm{z}\sim0.05$. Therefore each
absorber in the Pie diagrams was only paired with galaxies within a
very narrow redshift window, the size of which is too small to resolve
in Figure \ref{fig:piplots}.

\begin{figure*}
\begin{minipage}[t]{83mm}
\includegraphics[scale=0.45]{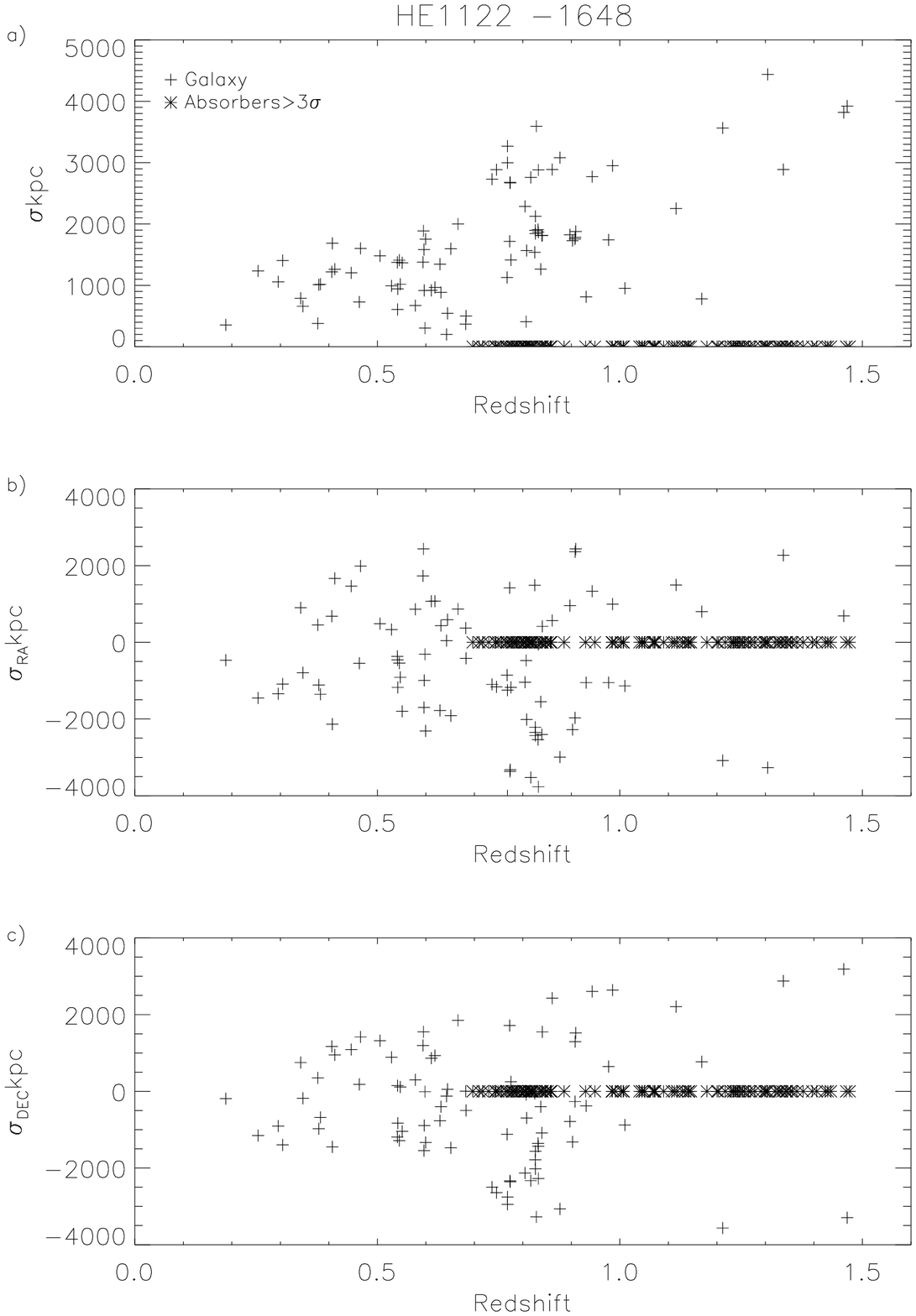}
\end{minipage}
\begin{minipage}[t]{83mm}
\includegraphics[scale=0.45]{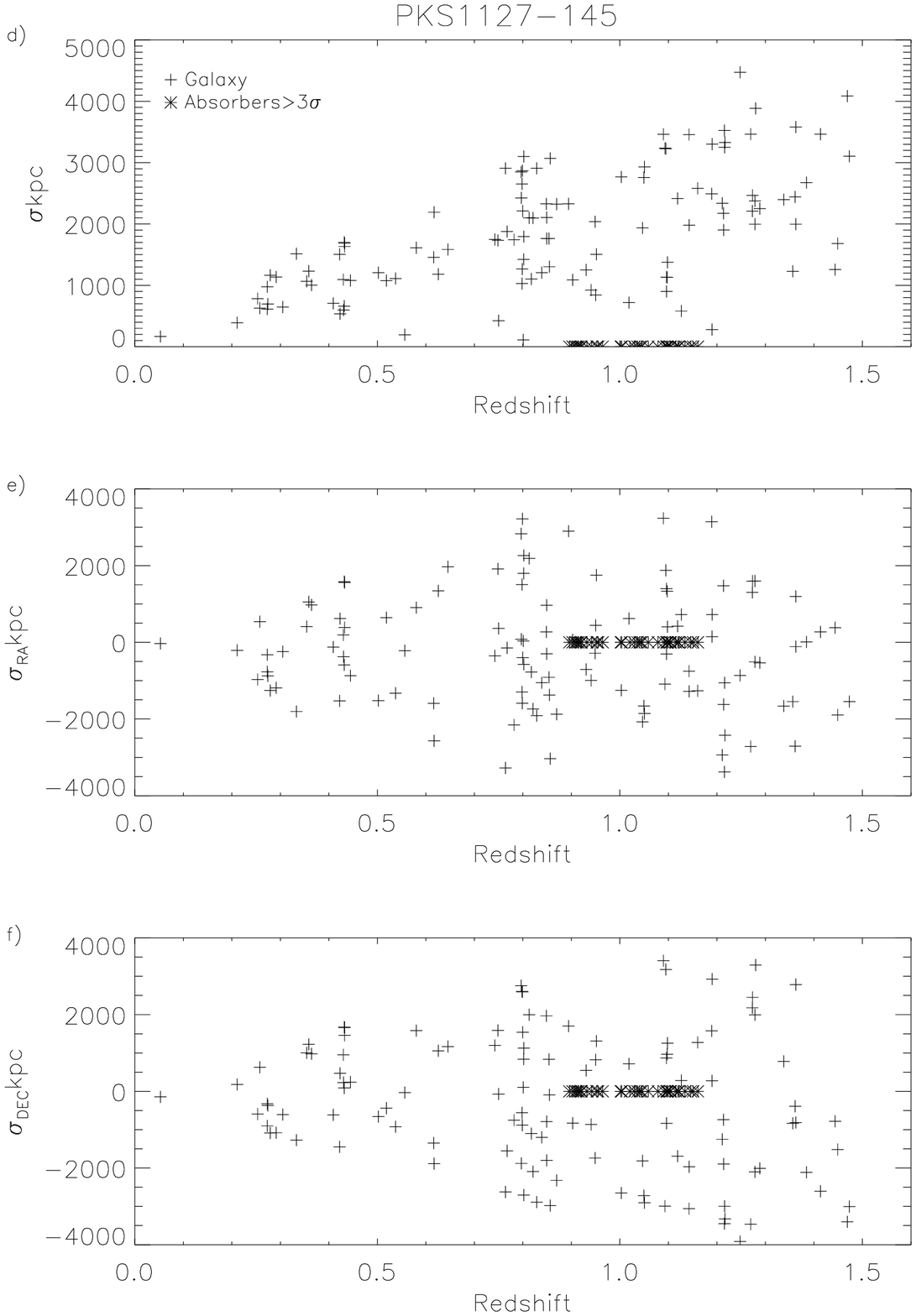}
\end{minipage}
\caption[Pie Diagrams for HE 1122-1648 and PKS 1127-145] {The
Pie-diagram for quasar fields HE 1122-1648 and PKS 1127-145. The top
panels, plots (a) and (d) show the impact parameter between the galaxy
and the quasar sightline plotted as a function of redshift. Panels
(b),(e), (c) and (f) show how this separation is split between the RA
and DEC of each quasar field respectively. The increasing maximum
separation with redshift, observed in plots (a) and (d) mark the
boundary of the survey volume.  The maximum redshift plotted only
covers the overlap between the two populations, so only absorbers from
the STIS E230M spectra have been included.  The narrower \lya\
absorber redshift window of the PKS 1127-145 field is because of the
lower redshift of this quasar.}
\label{fig:piplots}
\end{figure*}

\subsubsection{The Proximity Effect}

The line-of-sight proximity effect \citep{Bajtlik1988} describes the
process where ionising radiation from a quasar lowers the neutral
fraction of HI, thereby decreasing the line-density of \lya\
absorbers.  For lines close to the quasar redshift we may also detect
gas that is associated with the quasar itself.  This is a concern for
the PKS 1127-145 field since galaxies are located near to the QSO
redshift. For this reason the 3 \lya\ absorbers and 3 galaxies that were within 
3000 \kms\ \citep{ChenMulchaey2009} of the quasar were excluded
from the line-list and galaxy catalogue when correlating pairs.

\subsection{Results for the 2D 2-point Correlation Function}
Figure \ref{fig:absgalran}(a) shows the 145 real \lya\ absorber-galaxy
pairs out to $20\times5$ \hMpcsev\ binned with $\Delta\pi=2.0$,
$\Delta\sigma=1.0$ \hMpcsev. 
These pairs were made of \lya\ absorbers that had an equivalent width
significance of at least 3 and galaxies of a redshift confidence $\leq4$
according to the flags in Table \ref{tab:emabslines}.

Above $\sigma\geq4$ \hMpcsev\ there were
very few real pairs detected. For this reason this last column should
be ignored in subsequent plots as the Poisson noise introduced large
fluctuations in the results for $\xi_{\rm{AG}}$ in these bins.  Where
this occurred the highest value in the colour bar of all of the plots
ignores these maxima. Figure \ref{fig:absgalran}(b) shows the plot of
the normalised RR counts. The DD/RR-1 correlation plot is in Figure
\ref{fig:absgalran}(c) and the associated error in Figure
\ref{fig:absgalran}(d).

\begin{figure*}
\includegraphics[scale=0.70]{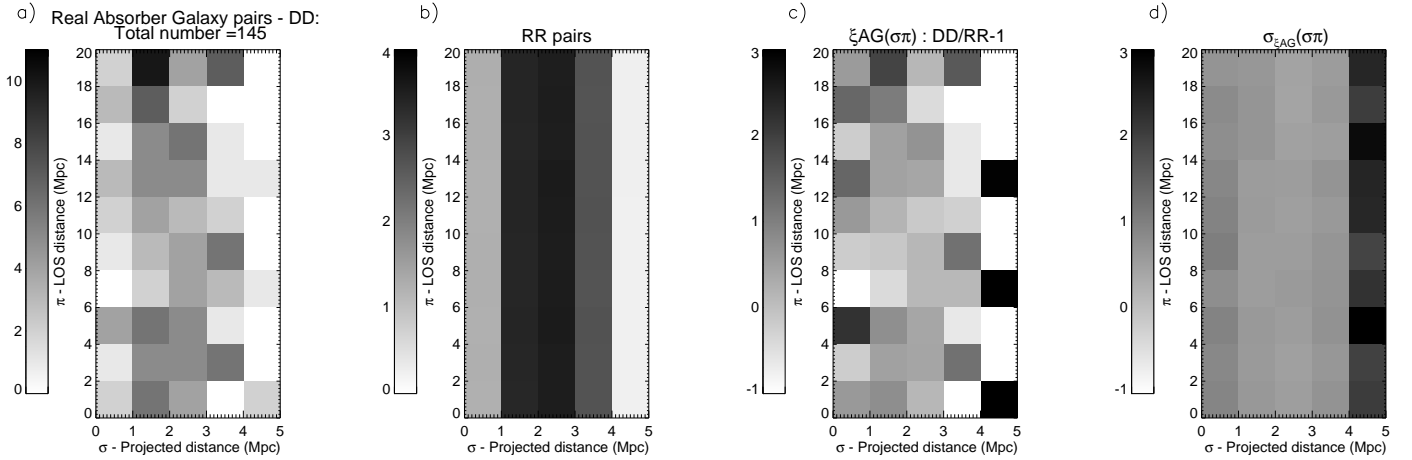}
\caption{Plot (a) shows the real absorber-galaxy pairs out to
$20\times5$ \hMpcsev, binned $\delta\pi=2.0\;\delta\sigma=1.0$
\hMpcsev. Those galaxies with a maximum confidence level of 4 have been included in this plot, 
with all \lya\ absorbers with
an equivalent width significance $\sigma_{\rm{EW}} \geq 3$.  The 2D
2-point correlation function $\xi_{\rm{AG}}=DD/RR-1$ was calculated
using 145 pairs. Plot (b) is of the random pairs normalised to the
same number of pairs as the real pair data.  Plots (c) and (d) show
$\xi_{\rm{AG}}$ and the associated error. There is no evidence for
correlation between galaxies and \lya\ absorbers within these
distances at this level of binning.  A $3\sigma$ upper-limit of
$\xi_{\rm{AG}}=2.8$ was computed using the bin with the maximum value
at $\pi=4-6\rm{,}\;\sigma=0-1$ \hMpcsev.  The central bin has a value
of $\xi_{\rm{AG}}=0.6\pm0.9$.}
\label{fig:absgalran}
\end{figure*}

Binned to this level we can conclude the following. 
\begin{itemize}
\item{For low column density absorbers (\logcoldens$\leq 17.4$ of which
only 11 have \logcoldens$\geq15$), there is
no evidence for correlation between the \lya\ absorbers and
galaxies. The maximum value in Figure \ref{fig:absgalran}(c) was
$\xi_{\rm{AG}}=2.2\pm 0.9$ at $\pi=4-6\rm{,}\;\sigma=0-1$
\hMpcsev. So within $20\times5$ \hMpcsev\ a $3\sigma$ upper-limit for
the cross-correlation is $\xi_{\rm{AG}}=2.8$.

The bin with the peak in the correlation function contained 4 real
pairs (1.25 random).  Excluding the $\sigma=4-5$ \hMpcsev\ separation 3 bins
contained no pairs giving $\xi_{\rm{AG}}=-1.0$.  One of these bins
occurred at $\pi=6-8$, $\sigma=0-1$ \hMpcsev\ with the minimum value of
$\xi_{\rm{AG}}=-1.0\pm 0.8$.  So there is a $2.6\sigma$ variation in
$\xi_{\rm{AG}}$ between the maximum and minimum values along the line
of sight at $\sigma=0-1$ \hMpcsev.  A similar variation in
$\xi_{\rm{AG}}$ is also shown at $\sigma=1-2$ \hMpcsev\ across the
same range in values. The values of $\xi_{\rm{AG}}$ varied from
$\xi_{\rm{AG}}=-0.4\pm 0.5$ to $\xi_{\rm{AG}}=1.9\pm 0.6$.
Comparable values of $\xi_{\rm{AG}}$ are also found at $\sigma=2-3$
\hMpcsev, with a peak of $\xi_{\rm{AG}}=0.7\pm 0.5$.  So there are no
significant trends observed in the degree of correlation along either
axis.}
\item{Binned to this level and at these redshifts obviously we see no
evidence for a ``finger-of-god'' at small projected separation along
the line-of-sight.}
\item{A result that was shared with Figure 2 of W07 is the suggestion
of an off-origin maximum in the correlation for low density absorbers
(\logcoldens$=13-15$), although this result was also consistent with a
correlation equal to 0.  The correlation function at the bin of
closest separation in Figure \ref{fig:absgalran}(c) is $\xi=0.6\pm
0.9$. This is only $1.2\sigma$ lower than the peak along $\sigma=0-1$
\hMpcsev.  That both investigations hint at a drop in correlation at
small separation is only tentative evidence that low-column density
\lya\ absorbers are not observed at short distances.  As their column
density is so low these absorbers could be easily photo-ionised and
stripped of their neutral gas when they enter the halo.}
\end{itemize}
We expect lines of a higher column density to show a greater
clustering. However, with such a small line-list and so few absorbers
that had \logcoldens$\ge15$ we have not measured the cross-correlation
of a sub-set with absorbers that have a higher column density.  Only
18 pairs fit this criteria with 2 real pairs in the bin of smallest
separation.

\subsection{Comparison with the Galaxy Auto-Correlation Function}
To compare with the cross-correlation, the galaxy auto-correlation was
calculated for all the galaxies with a redshift confidence $\leq4$ 
 that were above a redshift of
$\rm{z}=0.68$.  All literature results point to the galaxy
auto-correlation peaking at the
smallest separation then decaying in a distorted radial pattern
\citep{Pollo2005,Hawkins2003,Coil2006,Guzzo2008,Zheng2007}.

In the galaxy auto-correlation function the distribution of the random
pairs was made by perturbing each real galaxy redshift by a random
amount that varied between $\delta\rm{z}=\pm\;0.01-0.1$. These
boundaries were chosen so that the large-scale redshift distribution
was maintained but any galaxy pair that had existed would now no
longer appear within the $20\times5$ \hMpcsev\ window.

The plots in Figure \ref{fig:galpert} show the result of this approach.
Plot (a) of Figure \ref{fig:galpert} are  the 138 DD galaxy
pairs. Figure \ref{fig:galpert} plot (b) is of the RR pairs. 
Plots (c) and (d) show \acf\ and
the error, with a central peak of $\xi_{\rm{GG}}=10.7\pm1.4$.

\begin{figure*}
\includegraphics[scale=0.70]{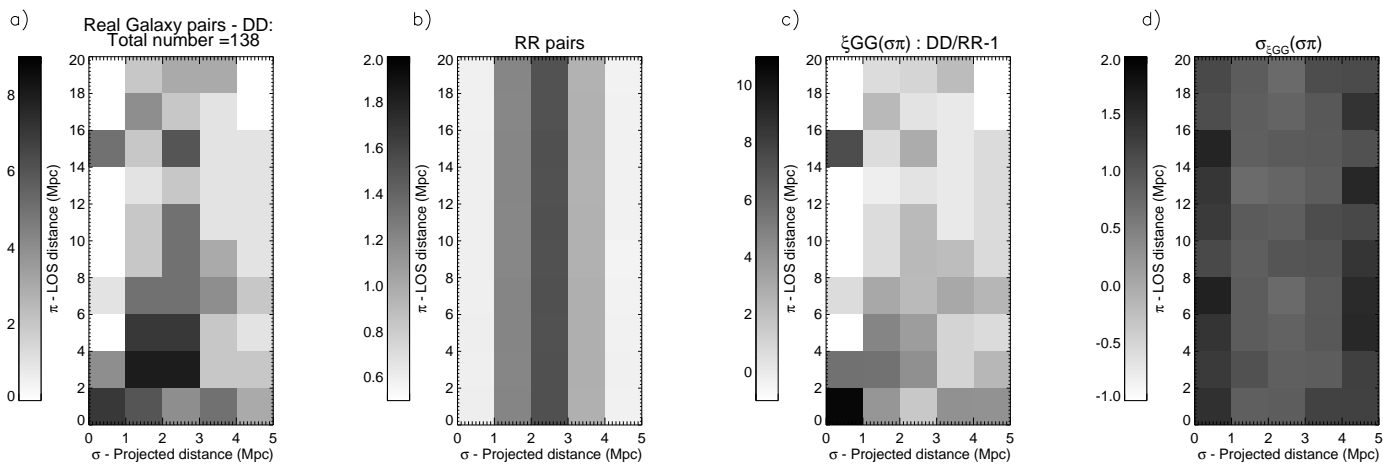}
\caption{Plots (a) and (b) show the DD and normalised RR pairs that were used for
  the galaxy auto-correlation function.  The redshifts for the
  RR pairs were created by perturbing each real redshift by
  $\delta\rm{z}=\pm\;0.01-0.1$. Plots (c) and (d) show
  $\xi_{\rm{GG}}\left(\sigma\rm{,}\pi\right)$ and the associated
  error. The peak in the auto-correlation in the bin of smallest separation
   is $\xi_{\rm{GG}}=10.7\pm1.4$.}
\label{fig:galpert}
\end{figure*}

The value in the central bin of  $\xi_{\rm{GG}}$
is in agreement with other galaxy auto-correlation measurements
taken at the same epoch, despite the smaller number of pairs in our sample. 
For example from the VVDS-Wide 
survey \citep{Pollo2005} which contained $\sim6000$ galaxy redshifts
at $0.6\leq\rm{z}\leq1.2$. The correlation function contained a peak of
$\xi_{\rm{GG}}\gtrsim10$ in the central bin and a prominent ``finger-of-god'' along
the line of sight \citep{Guzzo2008}.

It is interesting to study the evolution of the 
galaxy auto-correlation in order to compare this
with changes in the \lya\ absorber-galaxy cross-correlation.
The auto-correlation at small separation and at redshifts $\rm{z}\sim1$
is lower than at more recent epochs
because of the gravitational collapse of the halo systems \citep{Zheng2007}. 
For example Figure \ref{fig:Xihawk} shows
the auto-correlation of $\sim220000$ galaxies at redshifts  
$0.01\leq\rm{z}\leq0.2$ from the 2dFGRS data 
\citep{Hawkins2003}. These have has been re-binned to
match our plots so that the values for $\xi_{\rm{GG}}$ can be
compared.
 \footnote{Data supplied by Dr. P. Norberg, Institute for Astronomy,
  University of Edinburgh.} At redshits $\rm{z}\leq0.2$  
the value in the central
bin is $\xi_{\rm{GG}}=17.4\pm2.9$ and there is a significant ``finger-of-god'' 
along the line of sight. 

\begin{figure}
\begin{minipage}[h]{83mm}
\includegraphics[scale=0.60]{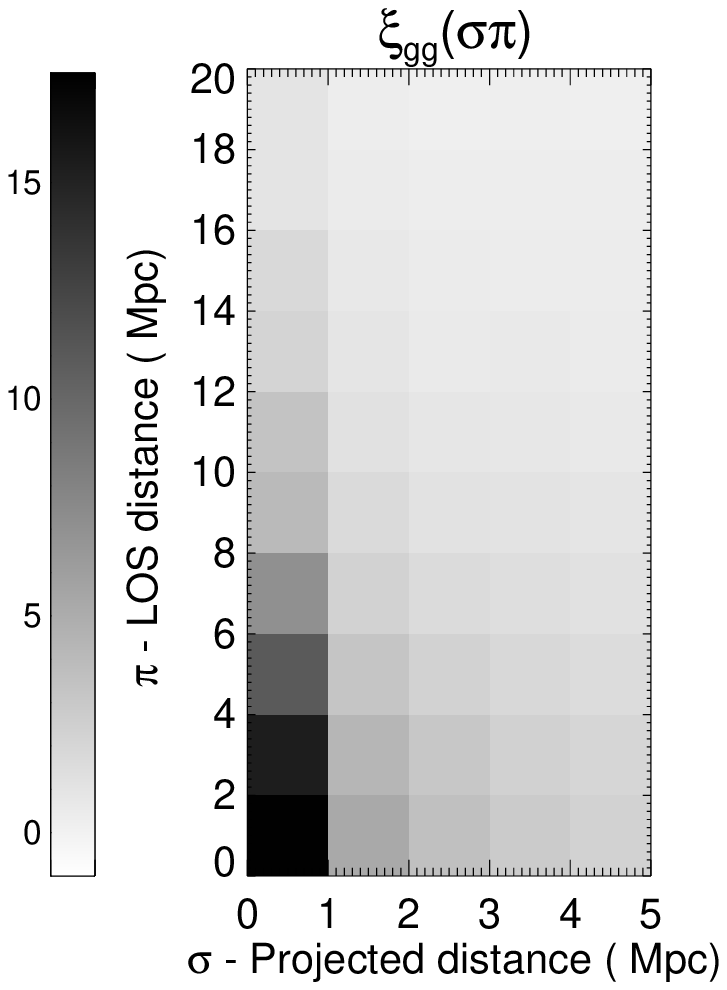}
\end{minipage}
\caption{The 2D 2-point auto-correlation $\xi_{\rm{GG}}$ of the 2dFGRS
from \citet{Hawkins2003}. The distances have been set to h=0.7 and the
data re-binned to $\delta\pi=2.0\rm{,}\;\delta\sigma=1.0$
\hMpcsev. When compared to figure \ref{fig:galpert} the values for
$\xi_{\rm{GG}}$ are  2.1 $\sigma$ higher at the centre with a
value of $\xi_{\rm{GG}}=17.4\pm2.9$.  We also see a significant
``finger-of-god'' redshift distortion along $\sigma=0-1$
\hMpcsev. An increase in $\xi_{\rm{GG}}$ is expected at lower redshifts
due to gravitational collapse.}
\label{fig:Xihawk}
\end{figure}

An alternative way to convey the difference between the results 
from \citet{Hawkins2003} and our
galaxy survey is to plot the redshift-space correlation parameter
$\xi_{\rm{GG}}\left(\rm{s}\right)$, where,
\begin{equation*}
\rm{s}=\sqrt{\left(\pi^2+\sigma^2\right)}
\end{equation*}
This is shown in Figure \ref{fig:sphrspace}.

\begin{figure}
\begin{minipage}[h]{83mm}\includegraphics[scale=0.50]{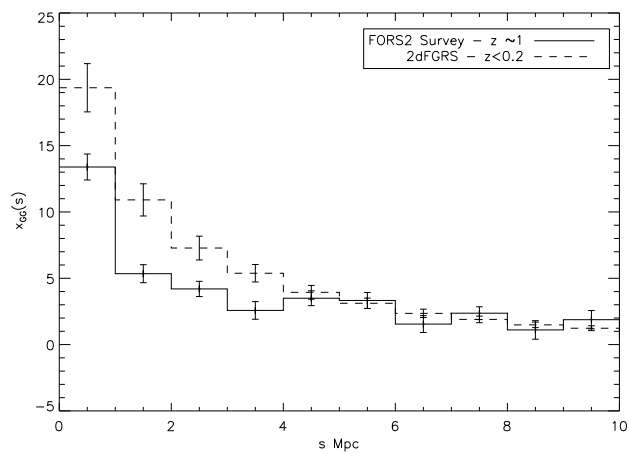}
\end{minipage}
\caption{This comparison between the auto-correlation from \citet{Hawkins2003} and
our result shows the $1.6\sigma$ increase in the auto-correlation  
$\xi_{\rm{GG}}\left(\rm{s}\right)$ of the central bin as the redshift has decreased
from $\rm{z}\sim1$ to $\leq0.2$.}
\label{fig:sphrspace}
\end{figure}

\subsubsection{Comparing the cross and auto-correlation function}
When $\xi_{\rm{AG}}$ from Figure \ref{fig:absgalran}(c) is compared with
$\xi_{\rm{GG}}$ from Figure \ref{fig:galpert}(c) we can draw the following
conclusion.  The values of the central bin have a
$5.9\sigma$ difference in the level of correlation, with a 
$5.4\sigma$ difference between the central peak of $\xi_{\rm{GG}}$ and
the $3\sigma$ upper-limit of $\xi_{\rm{AG}}$.

Therefore there is significant evidence of a difference in the cross-correlation
between galaxies and \lya\ absorbers and the galaxy auto-correlation
on scales of 1 Mpc at redshifts $\sim1$.

To check that our results are not dependent on galaxies with less certain 
redshifts, the cross and auto-correlation function were recalculated 
if pairs containing galaxies that had a redshift confidence of 4 were excluded. 
Galaxies that were identified only using [O II] emission constituted
the majority of the 95 galaxies of those that had a redshift greater than
$\rm{z}=0.68$.  This meant that there were too few pairs for both 
the cross and auto-correlation in each bin. In order to use the much
 smaller data sample we needed to increase the bin size.

\subsection{Variation in the binning and the 1D correlation function along $\pi$ and $\sigma$}
\subsubsection{Binned 2 $\times$ 2 \hMpcsev}
To reduce the effects of shot noise for the smaller galaxy sample 
we increased the size of each bin. The bin size
was increased to $\Delta\pi=\Delta\sigma=2$ \hMpcsev. However, 
the error in the cross-correlation was still comparable with the data 
in most bins. These new plots only extend to an impact parameter of 4 \hMpcsev.

Figures \ref{fig:abspert22} and \ref{fig:galpert22} show  $\xi_{\rm{AG}}$
and  $\xi_{\rm{GG}}$ respectively for those galaxies that had a redshift
confidence $\leq3$.

Figure \ref{fig:abspert22} supports the conclusion made earlier.  The
maximum value in the cross-correlation, not at the central bin, was now at
$\pi=12-14\rm{,}\;\sigma=0-2$ \hMpcsev\ with a value of $0.7\pm0.6$.
Therefore when binned to this extent there was only a weak 
correlation between \lya\ absorbers and galaxies. We found a $3\sigma$
upper-limit of $\xi_{\rm{AG}}=1.9$.

Comparisons with the galaxy auto-correlation also remained the
same. When binned to $\Delta\pi=\Delta\sigma=2$ \hMpcsev, (where averaging
over 2 \hMpcsev\ along $\sigma$ caused $\xi_{\rm{GG}}$ to decrease), in
the central bin there is still stronger clustering amongst galaxies
than between galaxies and \lya\ absorbers at a $6.0\sigma$ level of
confidence. The central bin in the auto-correlation now had a value of
\acf$=7.4\pm1.0$. 

\begin{figure*}
\begin{minipage}[t]{170mm}
\includegraphics[scale=0.70]{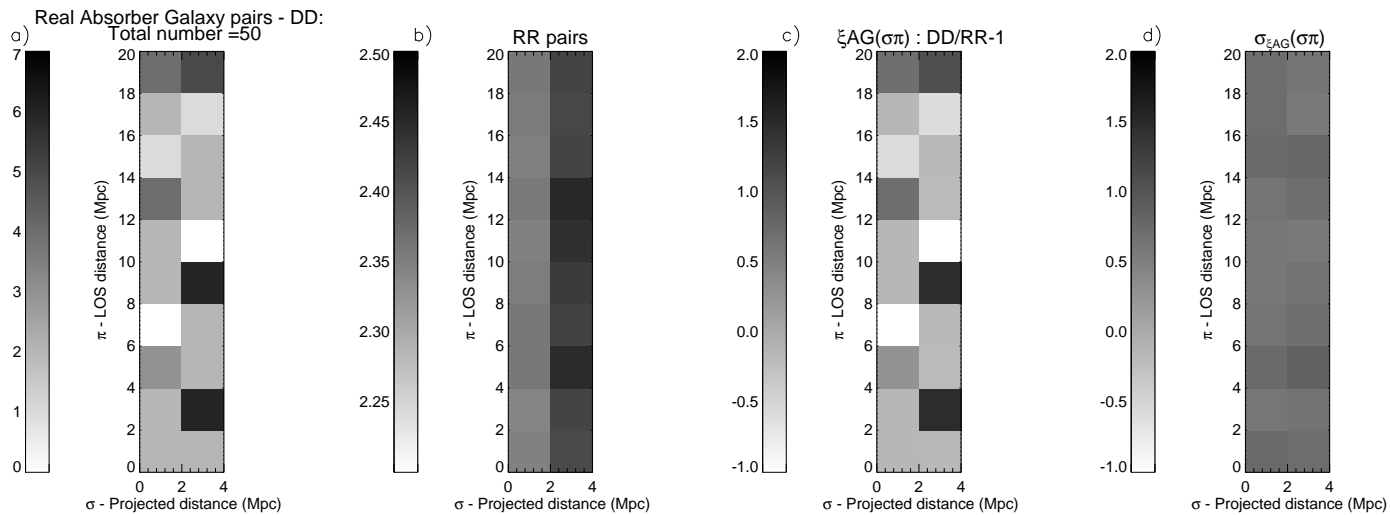}
\end{minipage}
\caption{Plots a-d are of the 50 absorber-galaxy DD pairs, RR
pairs, $\xi_{\rm{AG}}\left(\sigma\rm{,}\pi\right)$ and error
respectively when the level of binning has been increased to
$\Delta\pi=\Delta\sigma=2$ \hMpcsev.  The error per bin has decreased and
the peak in correlation is now $0.7\pm0.6$ at $\pi=12-14$, $\sigma=0-2$
\hMpcsev.  The result is the same as to what was observed in Figure
\ref{fig:absgalran}. This negligible degree of correlation
also does not change significantly along the line of sight as the
the value varies by $\sim1\sigma$ only.  Values for
$\xi_{\rm{AG}}$ also do not significantly change when at a projected
separation of $\sigma=2-4$ \hMpcsev.}
\label{fig:abspert22}
\end{figure*}

\begin{figure*}
\begin{minipage}[t]{170mm}
\includegraphics[scale=0.70]{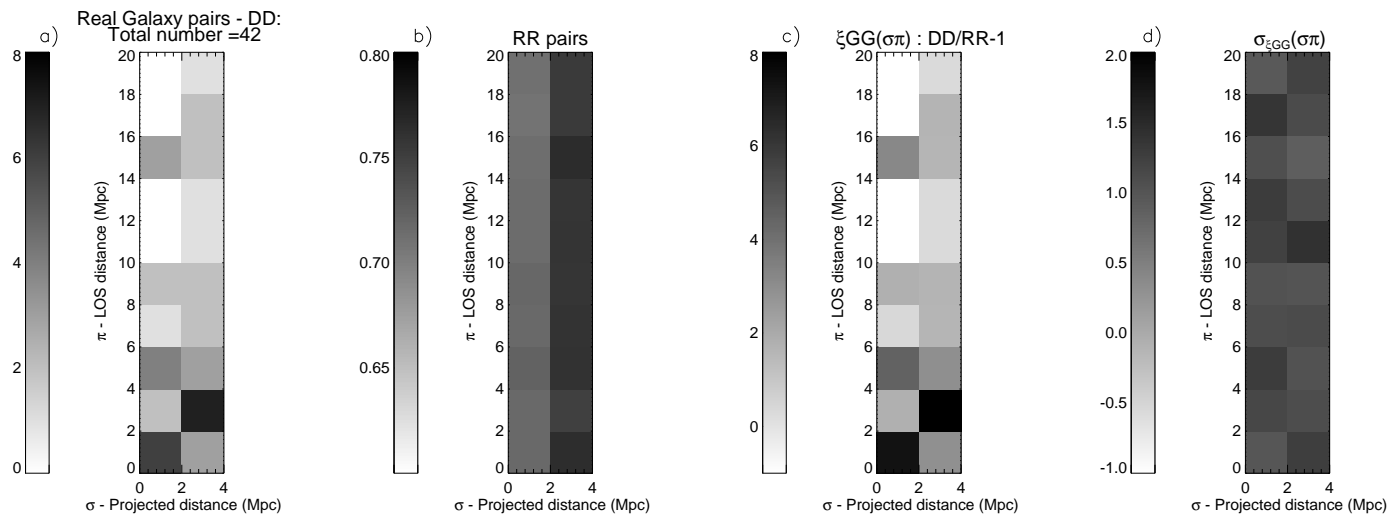}
\end{minipage}
\caption[$\xi_{\rm{GG}}\left(\sigma\rm{,}\pi\right)$ binned
$\Delta\pi=\Delta\sigma=2$ \hMpcsev.]  {Plots a-d are of the 42 real
galaxy pairs, DD, random-random galaxy pairs, RR, and the galaxy
auto-correlation, \acf\ and error respectively. When binned
$\Delta\pi=\Delta\sigma=2$ \hMpcsev\ the same conclusions can be drawn
for the auto-correlation function. Correlation between galaxies
decreased along the line of sight, until after $\pi\sim6$ \hMpcsev\ the
correlation signal was comparable with
$\xi_{\rm{AG}}\left(\sigma\rm{,}\pi\right)$.}
\label{fig:galpert22}
\end{figure*}

When the galaxies with a redshift confidence $\leq4$ were used
at this level of binning no significant difference was observed. The cross and
auto-correlation had values in the central bin of
$\xi_{\rm{AG}}=0.7\pm0.5$ and $\xi_{\rm{GG}}=6.1\pm0.7$ 
respectively.

\subsubsection{Summed over the line of sight and projected separation}
In order to further reduce the Poisson error in the correlation
function we collapsed the pair counts along each dimension
entirely. As in the previous section, in order to maximise the
number of pairs, we include galaxies with 
a redshift confidence $\leq$ 4. 

Plots (a), (b) and (c) of Figure \ref{fig:1Dabspert} show the number of
real pairs, the number of random pairs and $\xi_{\rm{AG}}$ when binned
along the projected separation with $\Delta\sigma=0.5$ \hMpcsev.  
This is the projected correlation $\Xi\left(\sigma\right)$ where;
\begin{equation*}
\Xi\left(\sigma\right) = \int\limits_{0}^{\infty}\xi(\sigma\rm{,}\pi)\;d\pi
\end{equation*}
In practice the upper limit was taken to be the maximum separation of
the pairs when binned in the 2D plots, $\pi=20$ \hMpcsev.  The absolute
separation between the absorbers and galaxies was used, therefore the
lower limit was zero. 

By summing over the line of sight we also removed any peculiar
velocity distortions that may have been present since these only
displace pairs along the $\pi$ axis.

The general shape of plot (a) of Figure \ref{fig:1Dabspert} supports
what has already been mentioned in Section \ref{sec:impactdist}.  As
the survey volume increased the number of galaxies, hence the number of pairs 
also increased. This peaked between 1-2 \hMpcsev\ before decreasing
because galaxies were now located at the edge of the surveyed
region. The smaller number of pairs at $\sigma=2-2.5$ \hMpcsev\ is real
and can be deduced from the Pie-diagrams in Figure \ref{fig:piplots}.  
In the field of view of HE 1122-1648 there
were only 4 galaxies that were at a separation of $2-2.5$ \hMpcsev. In the
field of PKS 1127-145 there were 20, however only 4 of these were
between the same redshift range as the absorbers.

The offset peak in plot (c) of Figure \ref{fig:1Dabspert} at
$\sigma=1\rightarrow 1.5$ \hMpcsev\ could be tentative evidence that
low column density absorbers are more scarce the closer the distance
to a galactic halo. However this peak is not significant with an
increase of only $1.3\sigma$ from $\sigma=0-0.5$ \hMpcsev. The level of
correlation then remained the same, as there was no significant change
in $\Xi_{\rm{AG}}\left(\sigma\right)$ out to $\sigma=4$ \hMpcsev.  The
2 bins above $\sigma=4$ \hMpcsev\ have been ignored in this comparison
as only 4 real pairs were found beyond this distance.

When binned using $\Delta\pi=2$ \hMpcsev\ along the line of sight, the
weak correlation $\xi_{\rm{AG}}\left(\pi\right)$ shown in plot (f) of Figure
\ref{fig:1Dabspert} did not have any significant change in value as
$\pi$ increased from $0-20$ \hMpcsev. 

\begin{figure*}
\includegraphics[scale=0.60]{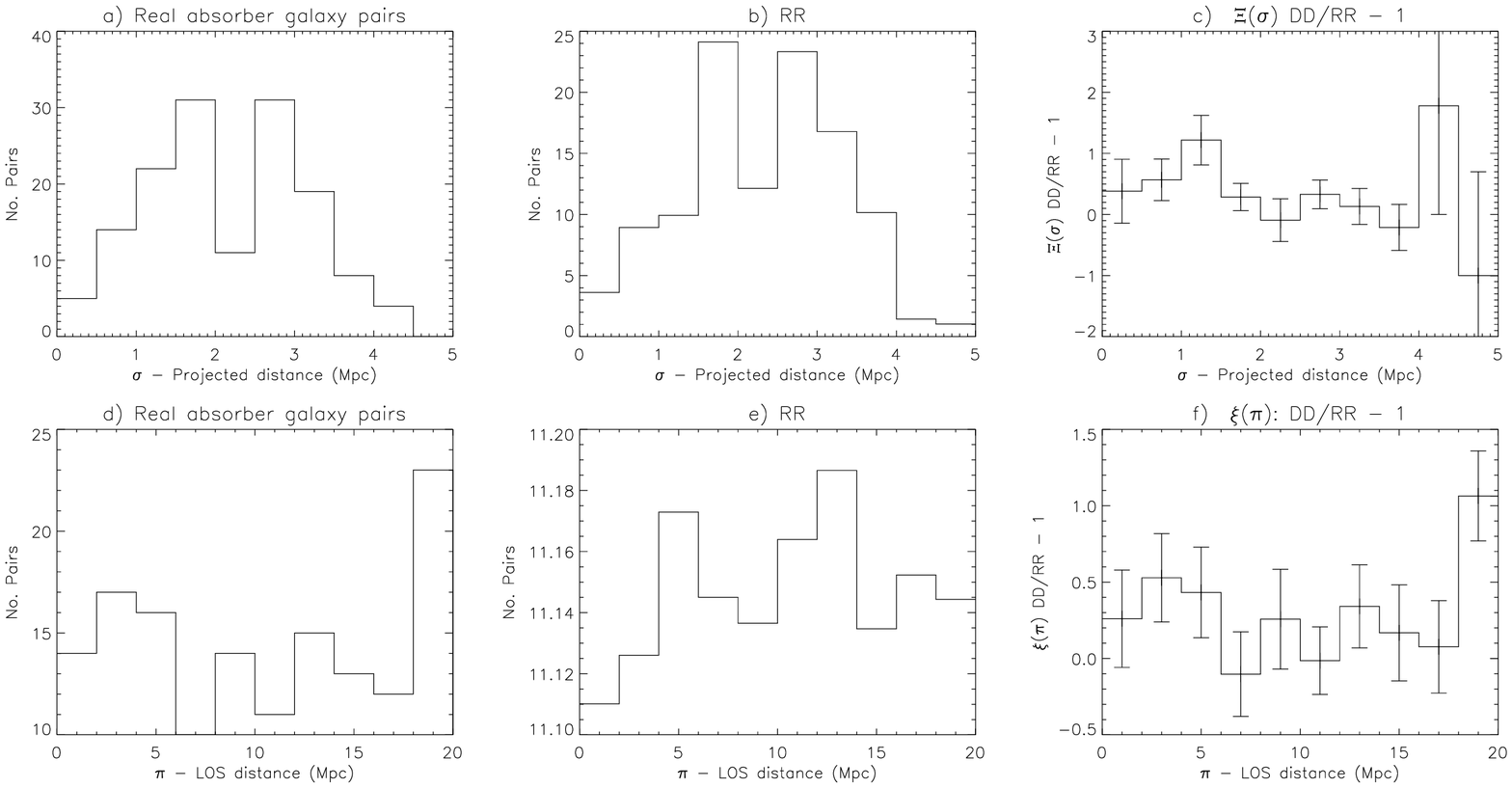}
\caption {Plots (a) and (b) show the number of DD and RR
absorber-galaxy pairs when they are binned along the projected
separation. The shape of these histograms are a consequence of the
edge effects discussed and illustrated in figure \ref{fig:velsep}. The
lack of pairs at $\sigma=2.5$ \hMpcsev\ is compensated in plot (b) by using
the perturbed redshift technique. Plot (c) suggests an off-centre peak
in the cross-correlation at $\sigma=1-1.5$ \hMpcsev. This peak is not
significant with an increase of only $1.3\sigma$ from $\sigma=0-0.5$
\hMpcsev. There is no significant change in
$\Xi_{\rm{AG}}\left(\sigma\right)$ out to $\sigma=4$ \hMpcsev.  Plots (d) and
(e) show the number of DD and RR absorber-galaxy pairs when binned along
the line of sight. Plot (f)
shows the cross-correlation, $\xi_{\rm{AG}}\left(\pi\right)$ in steps of 
$\delta\pi=2$ \hMpcsev. At these
levels of significance there is no evidence for any correlation between
the galaxies and low column density absorbers. What is important is
this lack of correlation is constant along the line of sight with a
variance of less that $1\sigma$ as $\pi$ increases from 0-18 \hMpcsev.}
\label{fig:1Dabspert}
\end{figure*}

\section{Discussion and Conclusion}
The two straw-man models mentioned in the introduction were that the
gas follows the galaxy distribution and collects in galaxy haloes, or
it forms part of a diffuse network, located near but not necessarily
causally linked with galaxies. 

Figure \ref{fig:absgalran} showed that we
did not detect significant correlation between galaxies  and \lya\
absorbers that have $13\leq$\logcoldens$\leq17.4$.  As only 145
absorber-galaxy pairs existed within
$\Delta\sigma=0-5$, $\Delta\pi=0-20$ \hMpcsev, in Figure
\ref{fig:absgalran}(a), no cuts in the column density were made as
this would further increase an already large uncertainty.  An
overwhelming majority of the \lya\ absorbers from the STIS spectra however
had \logcoldens$\leq15$.

We calculated a $3\sigma$ upper-limit of \twopcf$=2.8$. 
The decrease in the $3\sigma$ upperlimit to 1.9 in Figure
\ref{fig:abspert22}, caused by the increase in binsize to
$\Delta\sigma=\Delta\pi=2$ \hMpcsev\ is further evidence that \twopcf\
is near to 0 when averaged over these distances.  This result was
supported by the 1D plots of $\xi_{\rm AG}(\pi)$ and $\Xi_{\rm
AG}(\sigma)$ along the line of sight and projected separation in
Figure \ref{fig:1Dabspert}.

The value in the central bin for the galaxy auto-correlation function
was \acf$=10.7\pm1.4$. This was a $5.9\sigma$ increase on the value in
the equivalent bin of the absorber-galaxy cross-correlation that had a
value of \twopcf$=0.6\pm0.9$. 

Even when the value for the auto-correlation
was reduced by the larger binning in Figure \ref{fig:galpert22},
the difference between the 3-$\sigma$ upper-limit in the cross 
correlation and the central bin in the auto-correlation was still
$5.5\sigma$.

This evidence implies that at a redshift of $\rm{z}\sim1$ galaxies
that inhabit the nodes and filaments of the dark matter cosmic web are
not strongly correlated with the low column density \lya\ absorbers
(\logcoldens$\leq17$) that loosely trace this filamentary structure.

With a catalogue of 95 galaxies within a redshift range of
$\rm{z}=0.68-1.51$ that produced 138 galaxy pairs, it is unlikely that
we would have been able to reproduce the result 
and distribution of pairs of the galaxy auto-correlation function 
found by \citet{Pollo2005} and \citet{Guzzo2008}. However our results 
were similar within the margin of error.
When these studies are compared with the results at $\rm{z}\leq0.2$
there is a noticeable decrease in the auto-correlation.

When our results are compared with the values 
of the cross-correlation 
at redshifts $\lesssim0.5$ the data suggest that 
the cross-correlation between low column
density \lya\ absorbers and galaxies does not have a strong redshift
dependence. This may be because of the flat evolution in the 
\lya\ absorber density distribution at redshifts $\lesssim1.7$
\citep{Janknecht2006}.

This conclusion is a stark contrast to that which was found by RW06.
In that investigation \twopcf\ had been found to be comparable with
\acf\ with values $\approx11$ in the closest bin. There are three
possible reasons for this difference. First of all the dataset of RW06
was at low redshift, where clustering is thought to have increased
because of gravitational collapse. However an increase of this
magnitude over a change in redshift of $\rm{z}\sim1\rightarrow0$ does
not seem plausible, particularly when results gathered by
\citet{Wilman2007}, \citet{Chen2005,ChenMulchaey2009} 
and \citet{Pierleoni2008} cite lower values for 
\twopcf\ at redshifts of $\rm{z}\lesssim0.5$ and 0.

Another reason may be because of the galaxy population. RW06 used the
HIPASS galaxies that were detected using 21 cm emission and targeted
in a blind survey. Therefore as well as being gas rich it included
galaxies of any mass. Hence a strong correlation and velocity
dispersion along the line-of-sight were expected as gas falls through
the potential of the filamentary structure and is accreted directly by
the low-mass galaxies. This is the method of `cold-accretion' proposed
in \citet{Keres2005}.

Our magnitude limited galaxies (M$_B\sim-17\rightarrow-21$) were
collected in an incomplete survey at higher redshifts than that of the
HIPASS catalogue.  Consequently the galaxies for which we were able to
get spectroscopic data would have a higher mass than most of the
HIPASS candidates. \citet{Keres2005} claim that the dominant mode of
accretion for galaxies with a dark halo mass $\rm{M}\gtrsim10^{11.4}$
\Msolar\ is the more conventional method; where gas is shock heated to
the virial temperature of the halo before cooling and collapsing onto
the galaxy. This could mean less neutral gas is expected in the
immediate vicinity of the galaxies in our sample when compared to
those from RW06.

Another reason why the cross-correlation may seem larger in
\citet{Weber2006} is because of the smaller bin size of
$\Delta\sigma=\Delta\pi=0.1$ ${\rm h_{100}^{-1}}$Mpc that was used. We could argue
that the absorbers that are at $\sigma<0.1$ ${\rm h_{100}^{-1}}$Mpc
are contained within the galaxy halo and may even be virialised, so
\twopcf\ per bin is likely to substantially increase.

Alternatively these discrepancies could be due to the problem
suggested by
\citet{Pierleoni2008}. \citet{Weber2006,Wilman2007,Chen2005,ChenMulchaey2009} and to a
greater extent our dataset all either contain few pairs and/or 
few lines-of-sight. The appearance or absence of artifacts such as a
strong ``finger-of-god'' in the correlation pattern could be due
to the small size of the sample.

Unfortunately because we only had 13 galaxies in the catalogue that
were dominated by absorption features we were unable to test the
hypothesis of \citet{Chen2005} and \citet{ChenMulchaey2009} that
\twopcf\ is stronger for those galaxies that are dominated by
emission and comparable to the emission dominated galaxy 
auto-correlation function.  
However we still see a low correlation in the smallest bin, 
despite the abundance of emission dominated galaxies in our sample.

We are hoping to improve on these findings with more data collected
from three further STIS E230M lines-of-sight and galaxies from the
VIMOS instrument. This will enable us to correlate galaxies and \lya\
absorbers out to a projected separation of $\sigma\sim7$ \hMpcsev\ at
a redshift of $\rm{z}\sim1$.  We also plan to investigate the 2D
2-point correlation function using the Galaxies-Intergalactic Medium
Interaction Calculation (GIMIC)  SPH simulation \citep{Crain2009}.

\section{A Summary of the results on the 2 Point Correlation Function}
We have set the first limits on the 2D 2-point correlation function
between \lya\ absorbers of a column density $13.2\leq$\logcoldens$\leq17.4$ 
and galaxies from a magnitude limited survey of $21.5\leq R\leq24.5$  
at $\rm{z}\sim1$, and have compared this result to the galaxy 
auto-correlation.

When binned over both the projected separation and line of sight
out to 5 and 20 \hMpcsev\ respectively with
$\Delta\sigma=1$, $\Delta\pi=2$ \hMpcsev, there was no sign at a
3$\sigma$ level for any significant correlation. The peak value at
$\sigma=0-1$, $\pi=4-6$ \hMpcsev\ gave a $3\sigma$ upper-limit of
$\xi_{\rm{AG}}\left(\sigma\pi\right)=2.8$.
A $5.4\sigma$ difference was found between this upper-limit 
and our value for the galaxy auto-correlation function. 
This result is robust even when we remove subsets
of the galaxy sample that have less secure redshifts.

Our value for the galaxy auto-correlation, \acf$=10.7\pm1.4$ at $\sigma=0-1$, 
$\pi=0-2$ \hMpcsev\ was consistent with
the VVDS-Wide Survey that was taken at the same epoch.
Both of these surveys show a lower level of correlation than that
of the 2dFGRS survey at redshifts $\rm{z}\leq0.2$, which
is consistent with the evolution that is expected. 
 
We found no difference between our results and \twopcf\ found by
\citet{Wilman2007} who studied \twopcf\ at $\rm{z}\sim0.5$. Thus our
results are consistent with there being no significant evolution in
\twopcf\ with redshift. The much greater value of \twopcf\ found by
\citet{Weber2006} at $\rm{z=0}$ that was comparable to \acf\ could be
because the galaxies in her sample were gas rich and have a lower mass.

\section*{Acknowledgements}
Our thanks go to Robert Carswell for his support and
suggestions when using VPFIT.

Some of the data presented in this paper were obtained from the
Multimission Archive at the Space Telescope Science Institute
(MAST). STScI is operated by the Association of Universities for
Research in Astronomy, Inc., under NASA contract NAS5-26555. Support
for MAST for non-HST data is provided by the NASA Office of Space
Science via grant NAG5-7584 and by other grants and contracts.

The work in this paper was supported in part by an STFC PhD studentship.
 
\bibliography{paperbib}
\bibliographystyle{mn2e}
\bsp
\label{lastpage}
\end{document}